\documentclass[superscriptaddress,aps,prd,nofootinbib,reprint]{revtex4-2}

\usepackage{graphicx}
\usepackage{braket}
\usepackage[colorlinks,allcolors=blue]{hyperref} 
\usepackage{amsfonts,amsmath,amssymb,tensor}
\usepackage{Macro} 

\newcommand{\bR}[1]{{\mathbf{#1}}}
\newcommand{\al}[1]{\begin{align} #1 \end{align}}
\newcommand{\PD}[2]{\dfrac{\partial #1}{\partial #2}}
\newcommand{\Mat}[1]{\begin{pmatrix} #1 \end{pmatrix}}
\newcommand{\Tr}{\mathrm{Tr}}
\newcommand{\df}{\mathrm{d}}
\newcommand{\p}{\partial}
\newcommand{\fsky}{f_{\rm sky}}
\newcommand{\hC}{\widehat{C}}

\begin{document}

\title{
Is cosmic birefringence due to dark energy or dark matter? A tomographic approach
}

\author{Hiromasa Nakatsuka}
\affiliation{ICRR, University of Tokyo, Kashiwa, 277-8582, Japan}
\author{Toshiya Namikawa}
\affiliation{Kavli IPMU (WPI), UTIAS, University of Tokyo, Kashiwa, 277-8583, Japan}
\author{Eiichiro Komatsu}
\affiliation{Max Planck Institute for Astrophysics, Karl-Schwarzschild-Str. 1, 85748 Garching, Germany}
\affiliation{Kavli IPMU (WPI), UTIAS, University of Tokyo, Kashiwa, 277-8583, Japan}


\begin{abstract}
A pseudoscalar ``axionlike'' field, $\phi$, may explain the $3\sigma$ hint of cosmic birefringence observed in the $EB$ power spectrum of the cosmic microwave background (CMB) polarization data. Is $\phi$ dark energy or dark matter? A tomographic approach can answer this question. The effective mass of dark energy field responsible for the accelerated expansion of the Universe today must be smaller than $m_\phi\simeq 10^{-33}$~eV. If $m_\phi \agt 10^{-32}$~eV, $\phi$ starts evolving before the epoch of reionization and we should observe different amounts of birefringence from the $EB$ power spectrum at low ($l\alt 10$) and high multipoles. Such an observation, which requires a full-sky satellite mission, would rule out $\phi$ being dark energy. If $m_\phi \agt 10^{-28}$~eV, $\phi$ starts oscillating during the epoch of recombination, leaving a distinct signature in the $EB$ power spectrum at high multipoles, which can be measured precisely by ground-based CMB observations. Our tomographic approach relies on the shape of the $EB$ power spectrum and is less sensitive to miscalibration of polarization angles.
\end{abstract}

\maketitle


\section{Introduction}
\label{sec_Introduction}

A pseudoscalar ``axionlike'' field is a candidate for dark matter and dark energy in the Universe~\cite{Marsh:2015xka,Ferreira:2020fam}.
Like pion in the standard model of elementary particles and fields, a pseudoscalar can couple to the electromagnetic tensor $F_{\mu\nu}$ and its Hodge dual $\tilde F^{\mu\nu}$ via a Chern-Simons term in the Lagrangian density, ${\cal L}\supset -\frac14 g\phi F_{\mu\nu}\tilde F^{\mu\nu}$~\cite{Ni:1977,Turner:1987bw}, where $g$ is the axion-photon coupling constant.
This term rotates the plane of linear polarization of photons as they travel through space filled with $\phi$~\cite{Carroll:1989vb,Carroll:1991zs,Harari:1992ea}. 

Such a rotation produces non-zero odd-parity $TB$ and $EB$ power spectra of the cosmic microwave background (CMB) polarization fields, which vanish in the standard cosmological model~\cite{Lue:1998mq}.
This effect is often referred to as ``cosmic birefringence,'' as it resembles birefringence in a material (see Ref.~\cite{Komatsu:2022nvu} for a review).

The plane of linear polarization of CMB photons rotates clockwise on the sky by an angle 
$\beta=\frac12g\int_{t_\mathrm{LSS}}^{t_0}dt~d\phi/dt$, where $d\phi/dt$ is the total derivative of $\phi$ along the photon trajectory, and the subscripts ``0'' and ``LSS'' denote the present day and the last scattering surface of CMB photons, respectively. The CMB is an ideal target for measuring $\beta$, as it is proportional to the path length of photons when $\phi$ is evolving slowly.

Cosmic birefringence can be caused by $\phi$ of dark energy~\cite{Carroll:1998zi,Panda:2010uq} and dark matter~\cite{Finelli:2008,Fedderke:2019ajk}, as well as by possible signatures of quantum gravity~\cite{Myers:2003fd,Arvanitaki:2009fg}. How can we tell the origin? The effective mass of $\phi$, $m_\phi^2\equiv d^2V/d\phi^2$, is the key parameter, where $V(\phi)$ is the potential. The field does not change very much when $m_\phi\alt H(t)$, where $H(t)$ is the Hubble expansion rate at a time $t$; thus, $\phi$ would be dark energy today if $m_\phi\alt H_0\equiv H(t_0)\simeq 10^{-33}$~eV. The $\phi$ field with mass greater than this value would constitute a fraction of dark matter in the Universe today.

A tantalizing hint for $\beta$ has been
found in the $EB$ power spectrum of the \textit{Planck} mission with the statistical significance exceeding $3\sigma$~\cite{Minami:2020odp,Diego-Palazuelos:2022dsq,Eskilt:2022wav}. If confirmed with higher statistical significance in future, it would have profound implications for the fundamental physics behind dark energy and dark matter, as well as for quantum gravity. Anticipating such a discovery, in this paper we show how to determine $m_\phi$ using a tomographic approach to cosmic birefringence. 

There are two epochs in which linear polarization of the CMB was generated: (1) the epoch of recombination of hydrogen atoms and the subsequent decoupling of photons from plasma at a redshift of $z_\mathrm{rec}\simeq 1090$~\cite{Kosowsky:1994cy}; and (2) the epoch of reionization of hydrogen atoms at $z_\mathrm{rei}\simeq 7$~\cite{Zaldarriaga:1996ke}. The CMB photons that were last-scattered at these epochs would experience different amounts of cosmic birefringence~\cite{Liu:2006uh,WMAP:2008lyn}, which changes the relative amplitudes of the $EB$ power spectrum at low ($l\alt 10$) and high multipoles. We can use this $l$ dependence to infer $\beta$ from $z\simeq z_\mathrm{rec}$ to $z_\mathrm{rei}$ and that from $z\simeq z_\mathrm{rei}$ to $0$, i.e., tomography. For example, if $m_\phi \agt 10^{-31}$~eV, 
we should not detect the reionization bump
in the $EB$ spectrum at $l\alt 10$~\cite{Sherwin&Namikawa:2021}. 
Such an observation requires a full-sky satellite mission like \textit{LiteBIRD} \cite{LiteBIRD:2022} and would rule out $\phi$ being dark energy. 

The formula $\beta=\frac12g\int_{t_\mathrm{LSS}}^{t_0}dt~d\phi/dt$ assumes an instantaneous last scattering at $t_\mathrm{LSS}$, but a finite duration of last scattering leaves unique signatures in the CMB power spectrum~\cite{Finelli:2008}. If $m_\phi \agt 10^{-28}$~eV, $\phi$ starts oscillating during or earlier than the recombination epoch, modifying the $EB$ power spectrum at high $l$. This effect can be measured precisely by ground-based CMB observations such as Simons Observatory~\cite{SimonsObservatory:2018koc}, South Pole Observatory~\cite{Moncelsi:2020ppj}, and CMB-S4~\cite{CMB-S4:2016ple}, which opens up new scientific opportunities for CMB experiments.

In this paper, we solve the Boltzmann equation coupled with the equation of motion (EoM) for $\phi$, assuming that $\phi$ is either a dark energy field or a ``spectator'' field with negligible energy density. The energy density of $\phi$ therefore does not enter the Friedmann equation explicitly. We show how the shape of the $EB$ power spectrum depends on $m_\phi$, and provide a forecast for future constraints on
$m_\phi$ and $g\phi$.

The tomographic approach can also mitigate partially the artificial rotation angle, $\alpha$, by miscalibration of polarization angles of detectors and other instrumental effects~\cite{Miller:2009pt,QUaD:2008ado,WMAP:2010qai,Keating:2012ge,LiteBIRD:2021hlz}. The artificial rotation affects the $EB$ power spectrum at all multipoles equally, whereas the tomographic approach relies on the $l$-dependent effect. For example, the Galactic foreground emission experiences only a negligible amount of birefringence, and we can use the different $l$ dependence of the foreground and CMB power spectra to determine $\alpha$ and $\beta$ simultaneously~\cite{Minami:2019ruj}.
As we show in this paper, the difference between recombination and reionization signals can probe $10^{-32}~\mathrm{eV}\alt m_\phi\alt 10^{-31}$~eV (see Ref.~\cite{Sherwin&Namikawa:2021} for an earlier, more qualitative study), and details of the shape of the high-$l$ $EB$ power spectrum can probe $m_\phi \agt 10^{-28}$~eV.

The rest of the paper is organized as follows. In Sec.~\ref{sec_cmb_power_spectrum}, we present the Boltzmann equation and the EoM for $\phi$~\cite{Liu:2006uh,Finelli:2008,Gubitosi:2014cua,Lee:2016jym}. Our approach is different from Ref.~\cite{Cai:2021zbb}, which did not solve the EoM.
In Sec.~\ref{sec_isotropic_cosmic_birefringence}, we solve these equations to calculate the $EB$ power spectrum, and show new features that are important for cosmic birefringence tomography.
In Sec.~\ref{sec_experiment}, we forecast expected constraints on the axion parameters for experiments similar to \textit{LiteBIRD}~\cite{LiteBIRD:2022}, Simons Observatory~\cite{SimonsObservatory:2018koc}, and CMB-S4~\cite{CMB-S4:2016ple}.
We discuss possible improvements for our calculation in Sec.~\ref{sec_discussion} and conclude in Sec.~\ref{sec_conclusion}.

We use the Friedmann-Lema{\^\i}tre-Robertson-Walker spacetime with a metric tensor given by $a^2(\eta)\mathrm{diag}(-1,\mathbf{1})$, where $a(\eta)$ is the scale factor of the expansion of the Universe. We use the conformal time, $\eta$, as time coordinates unless noted otherwise.
We focus on the homogeneous axion background, $\phi(\eta)$, and ignore inhomogeneity in $\phi$.

\section{Boltzmann equation for isotropic cosmic birefringence}
\label{sec_cmb_power_spectrum}

\subsection{Setup}
\label{sec_setup}

We work with the Lagrangian density of axion electrodynamics given by \cite{Ni:1977,Turner:1987bw}
\begin{align}
    \mathcal L 
    =
    -\frac{1}{2} (\p_\mu \phi)^2
    -V(\phi)
    -\frac{1}{4} F_{\mu\nu} F^{\mu\nu}
    -\frac{1}{4} g\phi F_{\mu\nu} \tilde F^{\mu\nu} .
\end{align}
The dispersion relation of photons is given by
$    \omega_\pm^2 = 
    k^2
    \left(1\mp g \phi'/k \right)
    $,
where $\omega_\pm$ is the angular frequency of $\pm$ helicity states~\cite{Carroll:1989vb,Carroll:1991zs,Harari:1992ea}. The $+$ and $-$ states correspond to the right and left circular-polarization modes, respectively, in right-handed coordinates with the $z$ axis taken in the direction of propagation of photons. 
The prime denotes the derivative with respect to $\eta$.

In the WKB limit where $\phi$ varies slowly so that $\omega_\pm$ is much larger than the time evolution of $\phi$, i.e., $|\omega_{\pm}'|/\omega_\pm^2\ll 1$,
the rotation of the plane of linear polarization from $\eta$ to the present time
 is written as~\cite{Carroll:1989vb,Carroll:1991zs,Harari:1992ea}
\begin{align}
\label{eq:beta}
    \beta(\eta)
\equiv
    -
    \int^{\eta_0}_\eta \df \eta_1
    ~\frac{\omega_+ - \omega_-}{2}
=
    \frac{g}{2}\left[\phi(\eta_0) - \phi(\eta)\right] ,
\end{align}
where $\eta_0$ is the conformal time today. 
Here, we use the CMB convention for the position angle of linear polarization, i.e., $\beta>0$ is a clockwise rotation in the sky in right-handed coordinates with the $z$ axis taken in the direction of observer's lines of sight. The EoM for $\phi$ is
\begin{align}
	 \phi'' +2\frac{a'}{a}\phi' +a^2 m_\phi^2\phi = 0 ,
	\label{eq_EoM_phi}
\end{align}
for $V(\phi)=m_\phi^2\phi^2/2$.

The field does not evolve very much when $H(\eta)=a'/a^2\gg m_\phi$.
We choose the initial conditions at $\eta_\mathrm{in}$ such that $H(\eta_\mathrm{in}) \gg m_\phi$, $\phi'(\eta_\mathrm{in})=0$, and $\phi(\eta_\mathrm{in})=\phi_\mathrm{in}$.
We do not include the energy density of $\phi$ in the Friedmann equation explicitly, but use $H(\eta)$ derived from a flat $\Lambda$ cold dark matter ($\Lambda$CDM) model.
This approximation is valid when the energy density of $\phi$ is negligible (e.g., a tiny fraction of dark matter) or the axion mass is so small ($m_\phi\alt 10^{-33}$~eV) that it behaves as dark energy. The axion field with a tiny energy fraction can still induce a sizable amount of birefringence~\cite{Fujita:2020aqt}.

The EoM is a linear equation for $\phi$. We therefore introduce a function,  $f(\eta)\equiv\phi(\eta)/\phi_\mathrm{in}$, which  satisfies the same EoM as in \eq{eq_EoM_phi} with the initial condition $f(\eta_\mathrm{in})=1$. 
The birefringence angle is given by 
\begin{align}
\label{eq:beta2}
    \beta(\eta)
    = 
    \frac{g\phi_\mathrm{in}}{2} 
    \left[f(\eta_0)- f(\eta)\right]
    .
\end{align}

\subsection{Boltzmann equation}
\label{sec_boltzmann}

We work with parity eigenstates of CMB polarization, $E$ and $B$ modes, which have even and odd parity, respectively~\cite{Zaldarriaga:1996xe,Kamionkowski:1996ks}.
In the standard cosmological model, the $E$ and $B$ modes are uncorrelated due to parity symmetry. Cosmic birefringence violates parity symmetry and leads to a correlation between $E$ and $B$ modes~\cite{Lue:1998mq}.

In this paper, we consider only scalar-mode perturbations and ignore tensor modes. The evolution of linear polarization of CMB photons follows the Boltzmann equation~\cite{Kosowsky:1994cy}.
We expand Stokes parameters of linear polarization, $Q$ and $U$, in Fourier space with the wave vector $\mathbf{q}$. We define the cosine between $\mathbf{q}$ and the photon propagation direction as $\mu\equiv \mathbf{q}\cdot\mathbf{k}/(qk)$. We then write  the Boltzmann equation for the Fourier coefficients of $Q\pm iU$,  ${}_{\pm 2}\Delta_P(\eta, q, \mu)$, as~\cite{Komatsu:2022nvu}
\begin{align}
\nonumber
    {}_{\pm 2}\Delta_P' 
    +iq\mu~ {}_{\pm 2}\Delta_P
=&
    \tau'
    \left[
        -{}_{\pm 2}\Delta_P
        +\sqrt{\frac{6\pi}{5}}~ {}_{\pm 2}Y_l^0(\mu)\Pi
    \right]\\
&    \pm 2i\beta'~ {}_{\pm 2}\Delta_P
    , 
\end{align}
where ${}_{\pm 2}Y^m_l$ is the spin-2 spherical harmonics,
 $\Pi(\eta,q)$ is the polarization source term~\cite{Zaldarriaga:1996xe}, 
$\beta'\equiv g\phi'/2$ gives $\beta(\eta)=\int_\eta^{\eta_0}d\eta_1\beta'(\eta_1)$ in \eq{eq:beta}, and 
 $\tau'\equiv a(\eta)n_e(\eta)\sigma_T  $ is the differential optical depth with the Thomson scattering cross section $\sigma_T$ and the number density of electrons $n_e$.

With the $\mu$ dependence of ${}_{\pm 2}\Delta_P$ expanded in spin-2 spherical harmonics,
\begin{align}
    {}_{\pm 2}\Delta_P(\eta, q,\mu)
    &=
    \sum_l i^{-l} 
    \sqrt{4\pi(2l+1)}
    ~
    {}_{\pm 2}\Delta_{P,l}(\eta,q)~
    {}_{\pm 2}Y^0_l(\mu)
    ,
\end{align}
the formal solution for the Boltzmann equation is \cite{Liu:2006uh}
\begin{align}
\nonumber
    {}_{\pm 2}\Delta_{P,l}(\eta_0, q)
    =
    -
    \frac{3}{4} \sqrt{\frac{(l+2)!}{(l-2)!}} 
    \int^{\eta_0}_{0} & \df \eta~
    \tau' e^{-\tau(\eta)} 
    \Pi \frac{j_l(x)}{x^2} \\
    & \times \exp\left[\pm 2i\beta(\eta)\right]
    \label{eq_line_if_sight_DeltaP}
    ,
\end{align}
where $j_l(x)$ is the spherical Bessel function with $x=q(\eta_0-\eta)$ and 
$\tau(\eta)=  \int_\eta^{\eta_0}\df \eta_1\tau'(\eta_1) $.

Cosmic birefringence induces an imaginary part of ${}_{\pm 2}\Delta_{P,l}(\eta_0, q)$, which leads to $B$ modes. We write the coefficients of $E$ and $B$ modes as~\cite{Zaldarriaga:1996xe}
\begin{align}
    \Delta_{E,l}(q)
    \pm i  \Delta_{B,l}(q) 
    \equiv
    -{}_{\pm 2}\Delta_{P,l}(\eta_0, q)
    \label{eq_def_DeltaEB}
    .
\end{align}
Using Eqs.~\eqref{eq_line_if_sight_DeltaP} and \eqref{eq_def_DeltaEB}, the CMB polarization power spectrum is given by
\begin{align}
    C_l^{XY}
    =
    4\pi\int \df (\ln q)~
    \mathcal P_s(q)
    \Delta_{X,l}(q)
    \Delta_{Y,l}(q)
    ,
    \label{eq_CMB_power_spec}
\end{align}
where $\mathcal P_s(q)$ is the primordial scalar curvature power spectrum and $X,~Y=E$ or $B$. 
While cosmic birefringence modifies all polarization modes, we focus on the $EB$ power spectrum since it is more sensitive than $TB$ for the current and future generations of CMB experiments with low polarization noise.

We implement \eq{eq_line_if_sight_DeltaP} in the {\tt CLASS} code~\cite{Lesgourgues:2011re,blas2011cosmic} and calculate $C_l^{EB}$ with the best-fitting \textit{Planck} 2018 cosmological parameters for a flat $\Lambda$CDM model~\cite{Planck:2018vyg}. One significant change made to the code is the treatment of $B$ modes induced by scalar perturbations.
Cosmic birefringence transfers a part of scalar $E$ modes into $B$ modes, and we need to compute $\Delta_{B,l}$ which vanishes otherwise.

\subsection{Axion mass and the visibility function}
\label{sec_visibility_function}

\begin{figure}[t]
    \centering
    \includegraphics[width=\linewidth]{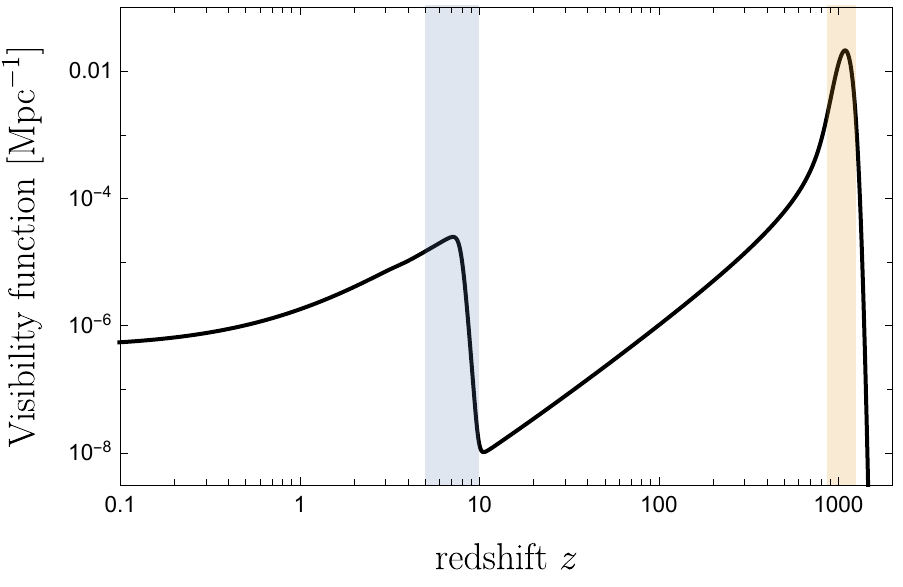}
    \caption{
    The visibility function as a function of redshift $z$.
    The orange and blue regions show the recombination and reionization epochs, respectively. The reionization is included using the \texttt{tanh} model with $z_\mathrm{rei} = 7.82$~\cite{Lewis:2008wr, Planck:2018vyg}.
    }
    \label{fig_visibilityfunction}
\end{figure}

Cosmic birefringence tomography relies on two epochs in which CMB polarization was generated. In Fig.~\ref{fig_visibilityfunction}, we show
the visibility function, the probability
density of photons being last scattered, defined by 
$    
    g_\mathrm{vis}(\eta) \equiv 
    \sigma_T a(\eta)n_e(\eta) e^{-\tau(\eta)}
$,
as a function of redshift with the thermal history obtained from the {\tt RECFAST} code~\cite{Seager:1999km,Seager:1999bc,Wong:2007ym}. 
As expected, the visibility function has the largest value at the recombination and photon decoupling epoch, $z_\mathrm{rec}\simeq 1090$. The second peak appears at $z_\mathrm{rei}\simeq 7$.

\begin{figure}[t]
    \centering
    \includegraphics[width=\linewidth]{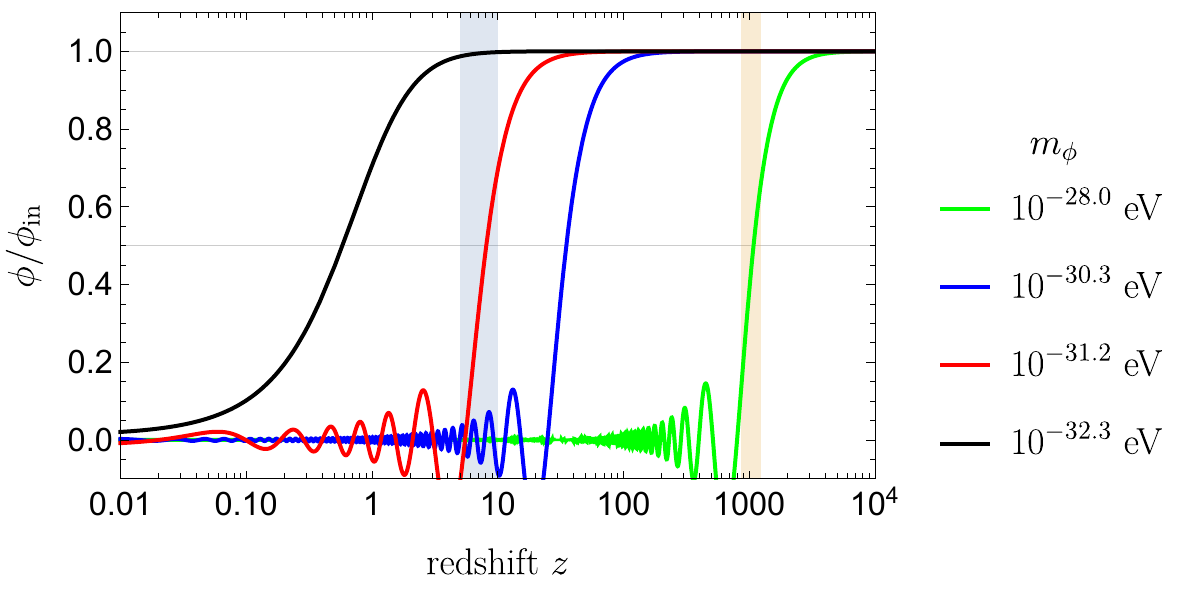}
    \caption{
    Evolution of $\phi$ for $m_\phi=10^{-28.0}$ (green), $10^{-30.3}$ (blue), $10^{-31.2}$ (red), and $10^{-32.3}$~eV (black). The shaded regions show recombination and reionization epochs as in Fig.~\ref{fig_visibilityfunction}.
    }
    \label{fig_field_dynamics}
\end{figure}

In Fig.~\ref{fig_field_dynamics}, we compare the evolution of $\phi$ with the epochs of recombination and reionization. 
The axion with $m_\phi=10^{-28.0}$~eV (green line) starts evolving significantly before recombination, which experiences a reduction in the amount of birefringence~\cite{Finelli:2008,Fedderke:2019ajk}.

The axion with $m_\phi=10^{-30.3}$~eV (blue line) starts evolving after recombination but has decayed before reionization; thus, very little cosmic birefringence would occur after reionization.
The axion with $m_\phi=10^{-31.2}$~eV (red line) starts oscillating during reionization, and that with $m_\phi=10^{-32.3}$~eV (black line) evolves only after reionization. The amplitude of the reionization bump in $C_l^{EB}$ is therefore sensitive to $10^{-32}~\mathrm{eV}\alt m_\phi\alt 10^{-31}$~eV.

For $m_\phi\alt 10^{-32}$~eV we expect $C_l^{EB}$ to be scaled by a single $\beta$ at all $l$. The polarization modes [\eq{eq_def_DeltaEB}] are simply given by
$    \Delta_{E,l}\pm i \Delta_{B,l} 
    = 
    e^{\pm 2i\beta}
    ( 
    \tilde \Delta_{E,l}\pm i \tilde \Delta_{B,l} 
    )
$,
where the tildes denote the values before cosmic birefringence. Then, the polarization power spectra after birefringence are given by~\cite{Feng:2004mq,Liu:2006uh} 
\begin{align}
    C_l^{EE} 
    &= \cos^2(2\beta) \tilde C_l^{ EE} + \sin^2(2\beta) \tilde C_l^{ BB} 
    ,
    \\
    C_l^{BB} 
    &= \cos^2(2\beta) \tilde C_l^{ BB} + \sin^2(2\beta) \tilde C_l^{ EE} 
    ,
    \\
    C_l^{EB} 
    &=
    \frac{1}{2}\sin(4\beta) (
    \tilde C_l^{ EE} - \tilde C_l^{ BB} 
    ) 
    .
    \label{eq_EB_const_betac}
\end{align}
In this case, $C_l^{EB}\propto C_l^{EE}$ when we ignore the primordial $B$ modes, and $\beta$ is degenerate with the instrumental miscalibration angle $\alpha$~\cite{QUaD:2008ado,WMAP:2010qai}. The tomography approach breaks this degeneracy by using the change in \textit{shape} of $C_l^{EB}$ induced by either $m_\phi\agt 10^{-32}$~eV~\cite{Sherwin&Namikawa:2021} or the Galactic foreground~\cite{Minami:2019ruj}.

\section{Cosmic birefringence from recombination and reionization}
\label{sec_isotropic_cosmic_birefringence}

\subsection{Toy example}
\label{sec_rotation_angle_step}

To build an intuitive understanding of the full numerical result for  $C_l^{EB}$ due to evolving $\phi$, we first study a toy example in which $\beta(z)$ integrated from $z$ to the present time [\eq{eq:beta}] changes abruptly:
\begin{align}
    	\beta(z) = 
    	\begin{cases}
    	        0&\quad\mathrm{for}\quad 
    	        z=0
    	        \\
    	        \beta_\mathrm{rei} 
    	        &\quad\mathrm{for}\quad 
    	        0<z\le 10
    	        \\
    	        \beta_\mathrm{rec}
    	        &\quad\mathrm{for}\quad 
    	        10<z
    	\end{cases}
    	,
		 \label{eq_beta_step}
\end{align}
where $\beta_\mathrm{rei}$ and $\beta_\mathrm{rec}$ are piecewise constant angles integrated out to $z=10$ and recombination, respectively.

\begin{figure}[t]
    \centering 
    \includegraphics[width=\linewidth]{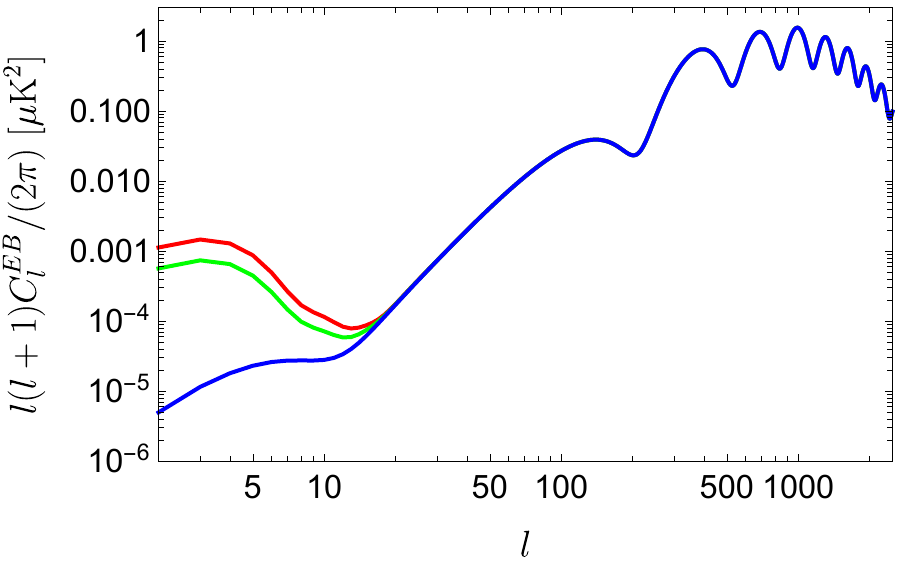}\\
    \includegraphics[width=\linewidth]{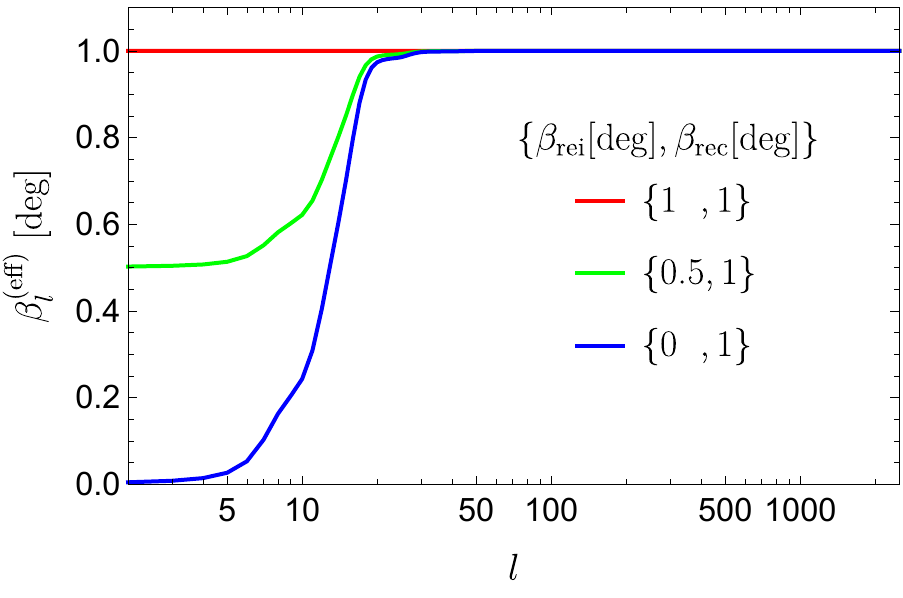}
    \caption{
    The $EB$ power spectrum from piecewise constant  angles given in \eq{eq_beta_step}: $\beta=\beta_\mathrm{rec}$ for $z>10$ and $\beta=\beta_\mathrm{rei}$ for $z\le 10$.
    The top and bottom panels show $l(l+1)C_l^{EB}/(2\pi)$ and the effective angles given in \eq{eq_betaeff}, respectively.
    }
    \label{fig_EB_for_stepalpha}
\end{figure}

In the top panel of Fig.~\ref{fig_EB_for_stepalpha} we show $C_l^{EB}$ for $\{\beta_\mathrm{rei}[\mathrm{deg}],~\beta_\mathrm{rec}[\mathrm{deg}]\} = \{1,~1\}$ (red), $\{0.5,~1\}$ (green), and $\{0,~1\}$ (blue). All lines coincide at $l\agt 20$ because $\beta_\mathrm{rec}$ is the same.
As $\beta_\mathrm{rei}$ decreases, the reionization bump of $C_l^{EB}$ also decreases. However, the reionization bump does not disappear even for $\beta_\mathrm{rei}=0$.

The shape of $C_l^{EB}$ can be understood as follows. The $E$ and $B$ modes are written as
\begin{align}
    \Delta_{E,l}\pm i \Delta_{B,l}
    &= 
    \sum_{x=\mathrm{rei,~rec}}
    e^{\pm 2i\beta_x}
    (\tilde \Delta_{E,l}^{(x)}\pm i \tilde \Delta_{B,l}^{(x)}).
\end{align}
Ignoring the primordial $B$ modes, $C_l^{EB}$ is given by
\begin{align}
\nonumber
    C_l^{EB} 
    =&
    \frac{1}{2}\sin(4\beta_\mathrm{rec}) \tilde C_l^{E\mathrm{rec},E\mathrm{rec}} 
    + \frac{1}{2}\sin(4\beta_\mathrm{rei}) \tilde C_l^{E\mathrm{rei},E\mathrm{rei}} \\&
    +\sin\left[2(\beta_\mathrm{rei}+\beta_\mathrm{rec})\right] \tilde C_l^{E\mathrm{rei},E\mathrm{rec}},
    \label{Eq:EB:rei-rec}
\end{align}
where $\tilde C_l^{Ex,Ey}$ is the cross power spectrum of $\tilde \Delta_{E,l}^{(x)}$ and $\tilde \Delta_{E,l}^{(y)}$ with $x,~y=$ rei, rec. 
The first term dominates at $l\agt 20$.
The second term produces the reionization bump at $l\alt 10$.
The third term was overlooked in Ref.~\cite{Sherwin&Namikawa:2021}.
The cross correlation of reionization and recombination $E$ modes induces a small reionization bump even when the rotation angle is zero at reionization.
This effect appears in the blue line of Fig.~\ref{fig_EB_for_stepalpha} where a small bump is seen at $l\alt 10$. Therefore, there is always some $C_l^{EB}$ at low $l$.

When the rotation angle depends on time, $C_l^{EB}$ is no longer given by \eq{eq_EB_const_betac}.
We thus define an effective angle at each $l$ as 
\begin{align}
    \beta_l^\mathrm{(eff)}
    \equiv
    \frac{1}{4} \arcsin\left( \frac{2x_l}{1+x_l^2}\right)
    \quad\mathrm{with}\quad
    {x_l = \frac{C_l^{EB}}{C_l^{EE}}},
    \label{eq_betaeff}
\end{align}
which is defined so as to reproduce \eq{eq_EB_const_betac} for the simplest case. Note that $\beta_l^\mathrm{(eff)}\simeq C_l^{EB}/(2C_l^{EE})$ for $C_l^{EB}/C_l^{EE}\ll 1$.
In the bottom panel of Fig.~\ref{fig_EB_for_stepalpha}, we show effective angles for the piecewise constant angles given in \eq{eq_beta_step}. 
The effective angle reproduces $\beta_l^\mathrm{(eff)} =\beta_\mathrm{rec}$ for $l\gg20$, where the recombination contribution dominates. It converges to $\beta_\mathrm{rei}$ for $l\ll20$, where the reionization contribution dominates. 

\subsection{\texorpdfstring{$EB$}{\textit{EB}} power spectrum from axion dynamics}
\label{sec_rotation_angle_axion}

\begin{figure*}[t]
    \centering 
    \includegraphics[width=.43\linewidth]{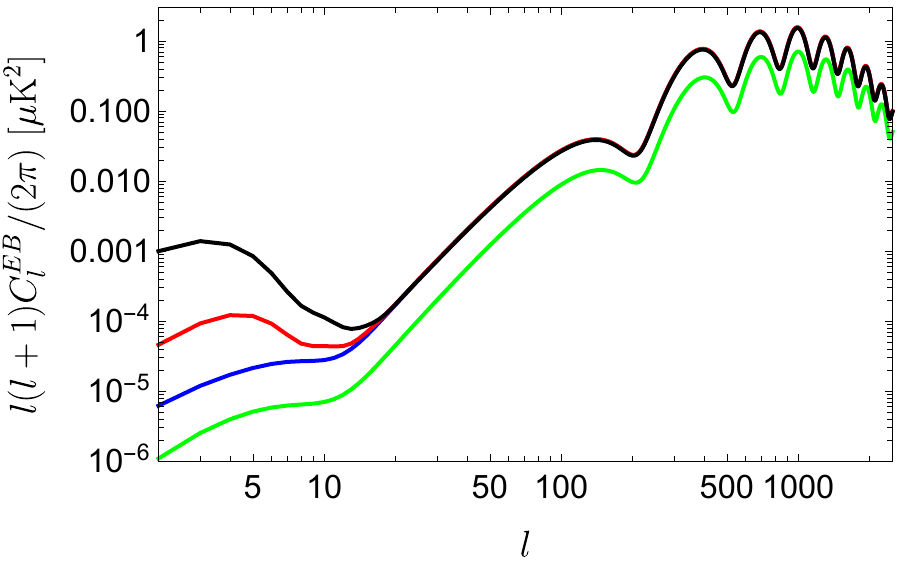}
    \includegraphics[width=.54\linewidth]{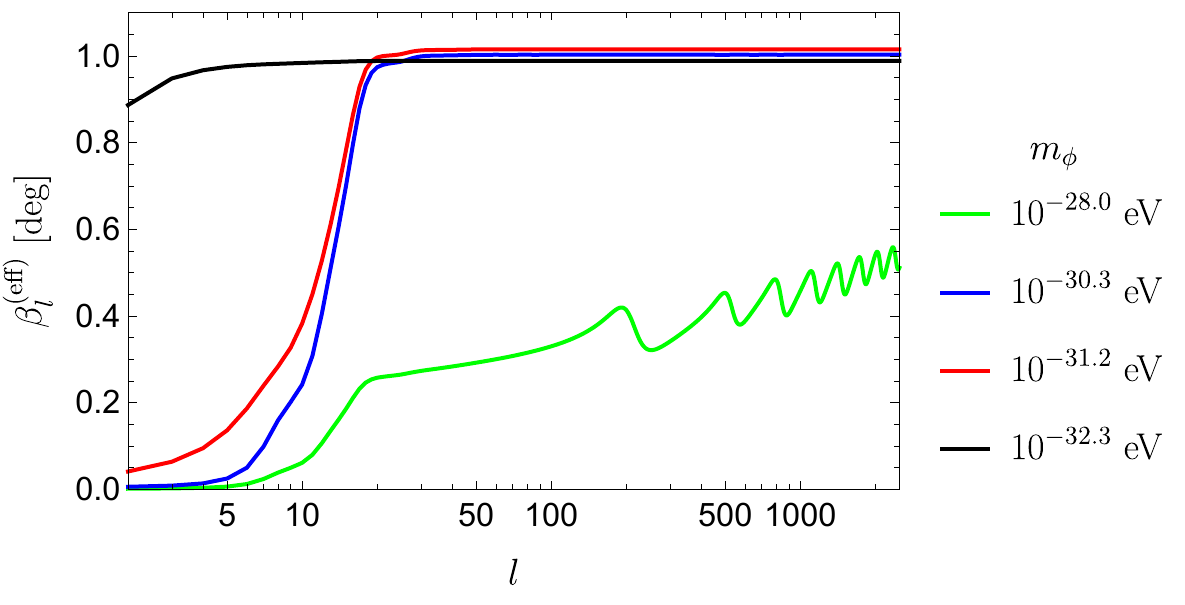}
    \caption{
Same as Fig.~\ref{fig_EB_for_stepalpha}, but for axion dynamics shown in Fig.~\ref{fig_field_dynamics}.
    }
    \label{fig_betal_for_mass}
\end{figure*}

In the left panel of Fig.~\ref{fig_betal_for_mass} we present the full Boltzmann solution for $C_l^{EB}$ with axion dynamics shown in Fig.~\ref{fig_field_dynamics}. In the right panel we show the corresponding $\beta^\mathrm{(eff)}_l$.
We use $g\phi_\mathrm{in}/2 = -1~\mathrm{deg} \simeq -0.01745$, which determines the overall amplitude of $\beta$ via \eq{eq:beta2}.

\subsubsection{
Reionization bump as a probe of \texorpdfstring{$m_\phi\alt 10^{-31}$~eV}{m < 1e-31~eV}
}
\label{sec:reionization}

We first study $m_\phi\alt 10^{-30}$~eV. As $\phi$ starts evolving well after the recombination epoch, the only difference appears in the reionization bump. 
We find the largest amplitude for $m_\phi\simeq 10^{-32}$~eV,
for which $\phi$ starts evolving only after reionization. The amplitude decreases as $m_\phi$ increases; however, the reionization bump does not disappear even for $m_\phi=10^{-30.3}$~eV, as explained in Sec.~\ref{sec_rotation_angle_step}.
We can therefore probe $m_\phi$ that falls between the black and blue lines in the left panel of Fig.~\ref{fig_betal_for_mass}.

\begin{figure}[t]
    \centering
    \includegraphics[width=\linewidth]{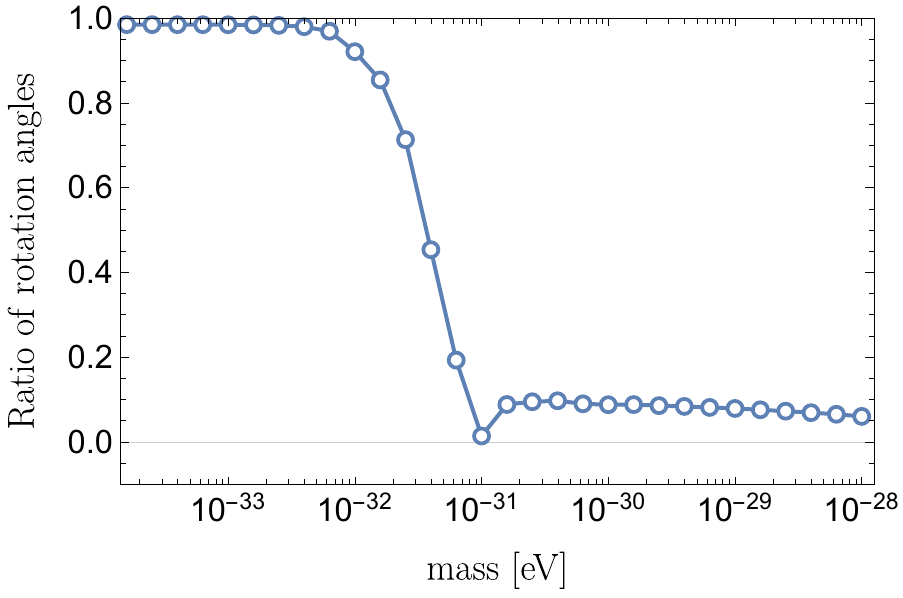}
    \caption{
    Ratio of the effective cosmic birefringence angles from reionization and recombination defined in \eq{eq_ratio_beta_def}. The ratio  is sensitive to the change in the axion mass over $10^{-32}~\mathrm{eV}\alt m_\phi\alt 10^{-31}$~eV.
    }
    \label{fig_betaratio}
\end{figure}

To make this statement more quantitative, we define the ratio of the effective angles for low and high $l$ as
\begin{align}
    \frac{\sum_{l=2}^{10}\beta^\mathrm{(eff)}_{l}/9}{
	\sum_{l=11}^{500}\beta^\mathrm{(eff)}_{l}/490
	}
	.
	\label{eq_ratio_beta_def}
\end{align}
In Fig.~\ref{fig_betaratio} we find that the
ratio is sensitive to the change in mass over $10^{-32}~\mathrm{eV}\alt m_\phi\alt 10^{-31}$~eV. We thus conclude that this is the range of $m_\phi$ we can probe using the relative amplitudes of the reionization bump and the high-$l$ $EB$ power spectrum.

\subsubsection{
\texorpdfstring{High-$l$}{High-l} features as a probe of \texorpdfstring{$m_\phi\agt 10^{-28}$~eV}{m > 1e-28~eV}
}
\label{sec_washout_recombination}

\begin{figure*}[t]
    \centering
    \includegraphics[width=.42\linewidth]{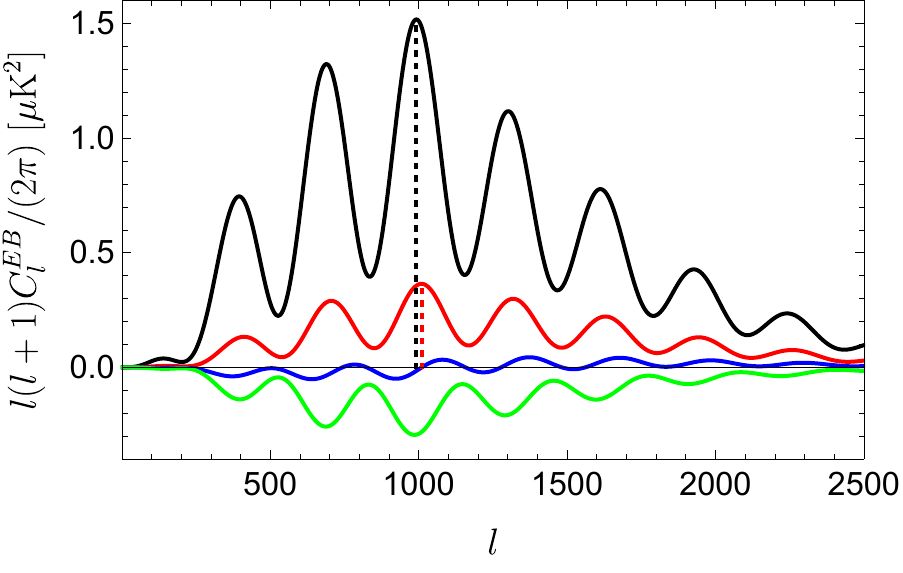}
    \includegraphics[width=.54\linewidth]{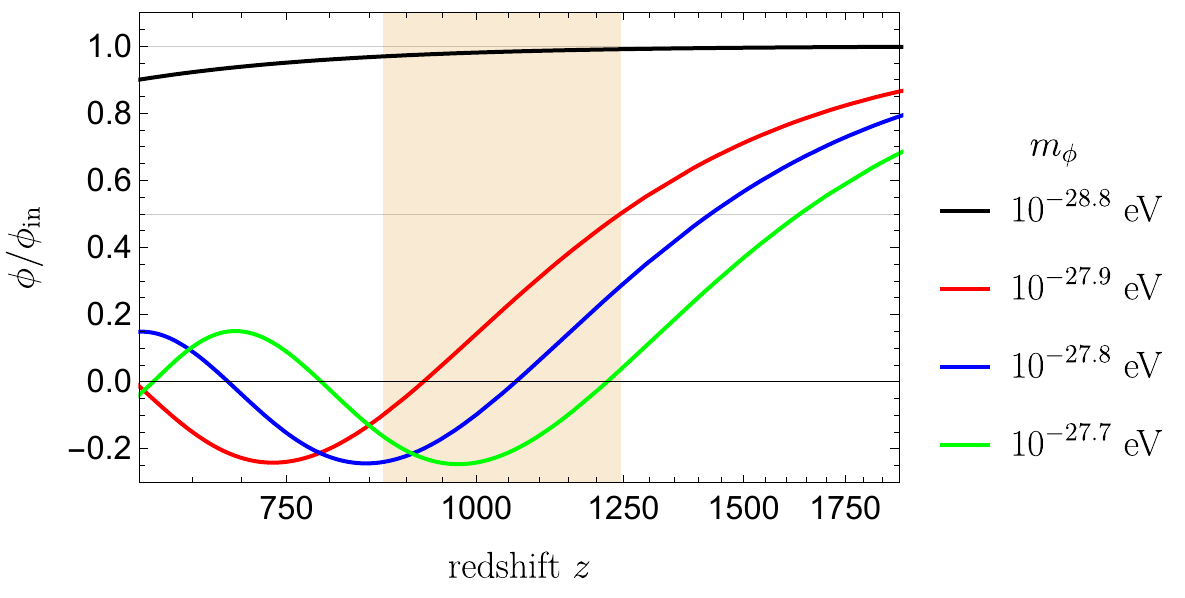}
    \caption{
    The $EB$ power spectrum for large $m_\phi$. The left panel shows $l(l+1)C_l^{EB}/(2\pi)$ for $m_\phi = 10^{-28.8}$ (black), $10^{-27.9}$ (red), $10^{-27.8}$ (blue), and $10^{-27.7}$~eV (green). The dotted vertical lines show the positions of the third peak. The right panel shows axion dynamics for each $m_\phi$. The shaded region shows the recombination epoch as in Fig.~\ref{fig_visibilityfunction}.
    }
    \label{fig_ClEB_heavies_axion}
\end{figure*}

The axion starts oscillating during recombination for $m_\phi \agt 10^{-28}$~eV. Photons last-scattered at different times experience different amounts of rotation of the plane of linear polarization, which can lead to partial cancellation  of cosmic birefringence~\cite{Finelli:2008,Capparelli:2019rtn,Fedderke:2019ajk} as well as to complex features in $C_l^{EB}$ at high $l$ that can be used to probe $m_\phi\agt 10^{-28}$~eV in a completely new way.

In the left panel of Fig.~\ref{fig_ClEB_heavies_axion} we show $C_l^{EB}$ for axion dynamics shown in the right panel. 
There are two effects on $C_l^{EB}$: the overall amplitude and shape.

We first discuss the amplitude, which changes dramatically depending on the value of $\phi$ during recombination.
For $m_\phi = 10^{-28.8}$~eV, $\phi$ is nearly constant, which results in $C_l^{EB}\propto C_l^{EE}$ except for the reionization bump.
For $m_\phi = 10^{-27.9}$~eV, $\phi$ starts evolving during recombination, resulting in a smaller $C_l^{EB}$.
For $m_\phi = 10^{-27.8}$~eV, $\phi$ averaged over recombination is tiny, resulting in a highly suppressed $C_l^{EB}$.
For $m_\phi = 10^{-27.7}$~eV, $\phi$ averaged over recombination is \textit{negative}, hence $C_l^{EB}<0$.

In the previous work that did not solve the Boltzmann equation, the amplitude of $C_l^{EB}$ has been calculated by averaging $\phi$ over the visibility function~\cite{Fedderke:2019ajk,Capparelli:2019rtn,Fujita:2020aqt}:
\begin{align}
    \braket{\beta}
    &\equiv \frac{g}{2}\left[\phi(\eta_0)-\braket{\phi}_\mathrm{LSS}\right],
    \label{eq_beta_rec_approx}
\end{align}
where
\begin{align}
\label{eq:phiLSS}
    \braket{\phi}_\mathrm{LSS}
    &\equiv
    \frac{
        \int^{\eta_{z=10}}_0 \df \eta ~g_\mathrm{vis}(\eta) \phi(\eta)
    }{
        \int^{\eta_{z=10}}_0 \df \eta ~g_\mathrm{vis}(\eta)
    }
    ,
\end{align}
with $\eta_{z=10}$ being a conformal time at $z=10$. Since we focus on the rotation angle from recombination, the average is limited to $z>10$ with $\int^{\eta_{z=10}}_0 \df \eta ~g_\mathrm{vis}(\eta)\simeq 0.95$. We use $g_\mathrm{vis}(\eta)$ computed with \texttt{CLASS} as shown in Fig.~\ref{fig_visibilityfunction}.

\begin{figure}[t]
    \centering
    \includegraphics[width=\linewidth]{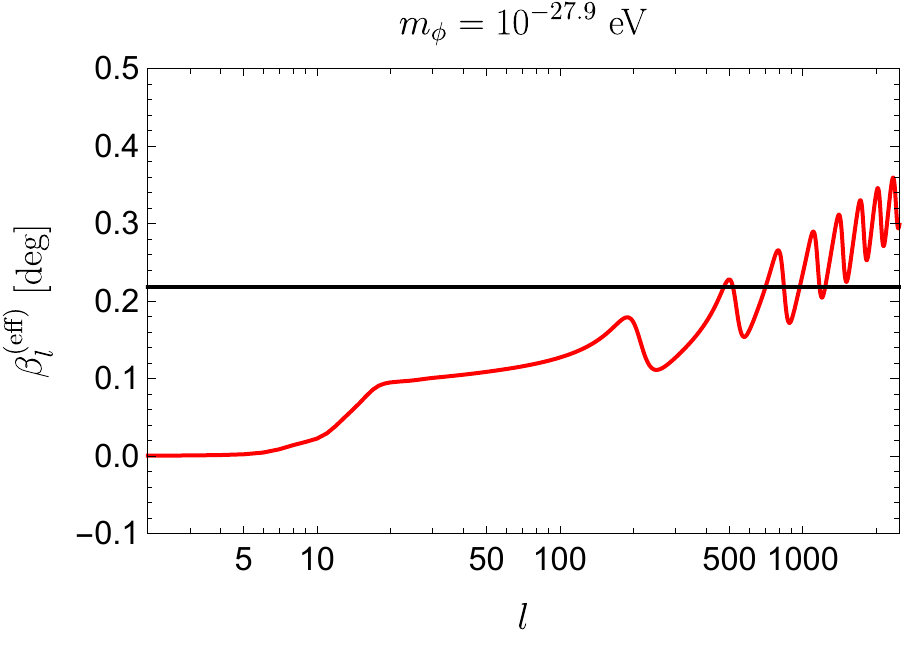}\\
    \includegraphics[width=\linewidth]{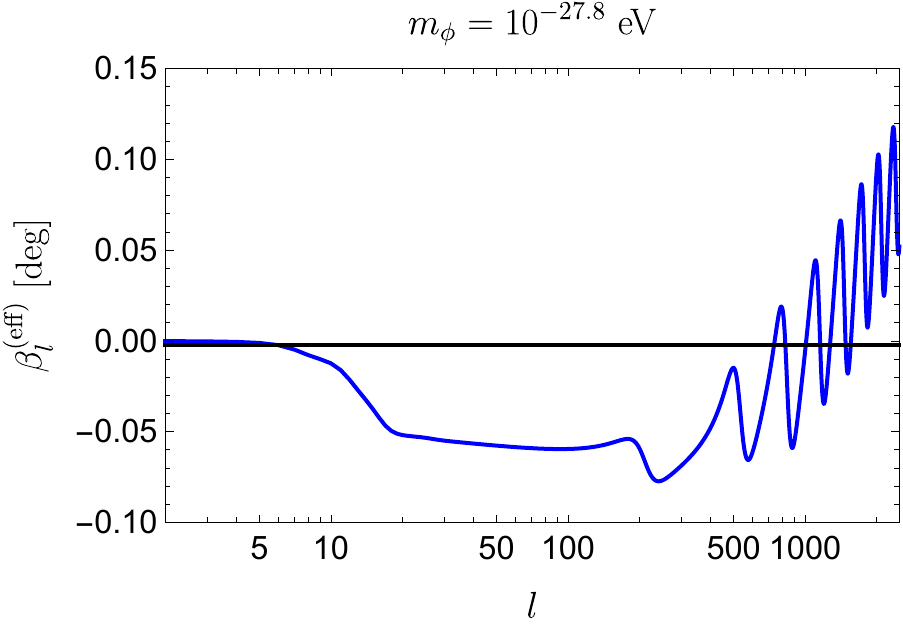}
    \caption{
    The effective rotation angles from the Boltzmann equation (wiggly lines) and $\braket{\beta}$ given in \eq{eq_beta_rec_approx} (horizontal lines) for $m_\phi = 10^{-27.9}$~eV (top) and  $10^{-27.8}$~eV (bottom).
    }
    \label{fig_betal_m27978_compare}
\end{figure}

In Fig.~\ref{fig_betal_m27978_compare} we compare $\beta^\mathrm{(eff)}_l$ computed from the Boltzmann equation and \eq{eq_beta_rec_approx} for $m_\phi=10^{-27.9}$~eV (top panel) and $m_\phi=10^{-27.8}$~eV (bottom). It is clear that $\beta^\mathrm{(eff)}_l$ shows much more complex features than just the average value $\braket{\beta}$ shown by the horizontal lines.

How can we understand such complex dependence of $\beta^\mathrm{(eff)}_l$ (hence $C_l^{EB}$) on $m_\phi$? We find that the location of the acoustic peaks in $C_l^{EB}$ for $m_\phi = 10^{-27.9}$~eV shifts to higher $l$ compared to that for $m_\phi = 10^{-28.8}$~eV (see the vertical dotted lines in Fig.~\ref{fig_ClEB_heavies_axion}).
This peak shift is the origin of the oscillating behavior of $\beta^\mathrm{(eff)}_l$ at $l>100$.

The peak location is determined by $D_A/r_s$, where $r_s$ and $D_A$ are the sound horizon and angular diameter distance at last scattering, respectively. In our fiducial cosmological model, $D_A/r_s\simeq (1+z_{\rm LSS})^{1/2}$ where $z_{\rm LSS}$ is the redshift of last scattering.
For $m_\phi = 10^{-27.9}$~eV, $\phi$ starts evolving before recombination, and the $B$ modes are mainly generated in the early stage of recombination.
In such case, $D_A/r_s$ for the induced $B$ modes becomes effectively large and the peaks shift to higher $l$.
If $m_\phi = 10^{-27.7}$~eV, the time when polarization is mostly produced is close to that for $m_\phi=10^{-28.8}$~eV and the peak locations are almost identical to those of the black line while the amplitude is negative. 

These features are important: we can use this complex dependence of $C_l^{EB}$ to determine $m_\phi$ in a completely new manner. While the overall amplitude $\braket{\beta}$ is degenerate with the miscalibration angle $\alpha$ (unless we have access to $l\alt 10$), the $l$ dependence is not.
This is relevant for ground-based CMB experiments (Sec.~\ref{sec:CMB-S4}).

\section{Forecast}
\label{sec_experiment}

\subsection{Simultaneous determination of \texorpdfstring{$\alpha$}{alpha} and cosmic birefringence with the reionization signal}
\label{sec:litebirdlike}

We first consider the case in which we simultaneously constrain cosmic birefringence and miscalibration angles, $\alpha$, using the reionization bump in $C_l^{EB}$~\cite{Sherwin&Namikawa:2021}. As explained in Sec.~\ref{sec:reionization}, the axion with $m_\phi\agt 10^{-32}$~eV changes relative amplitudes of $C_l^{EB}$ at low and high $l$, which cannot be mimicked fully by $\alpha$ because it affects all multipoles equally via \eq{eq_EB_const_betac} with $\beta\to \alpha$.

In Ref.~\cite{Sherwin&Namikawa:2021}, $C_l^{EB}$ was modeled as the sum of the reionization and recombination contributions: 
\begin{align}
    C^{EB}_l=\sum_{x={\rm rei,rec}}2\beta_x (C^{EE,x,{\rm lens}}_l-C^{BB,x,{\rm lens}}_l)
    \,, 
\end{align}
where $C^{EE,x,{\rm lens}}_l$ and $C^{BB,x,{\rm lens}}_l$ are the lensed $E$- and $B$-mode power spectra, respectively. 
As shown in \eq{Eq:EB:rei-rec}, however, we cannot decompose $C_l^{EB}$ in this way. 
Therefore, we constrain the axion parameters instead of $\beta_x$.

Assuming that the observed $E$- and $B$ modes obey a multivariate Gaussian distribution with zero mean, the Fisher information matrix is given by \citep{Tegmark:1996:Fisher}
\begin{align}
    \bR{F}_{ij} = \fsky\sum_{l=2}^{l_\mathrm{max}} \frac{2l+1}{2}
		\Tr\left(\bR{C}^{-1}_l\PD{\bR{C}_l}{p_i} \bR{C}^{-1}_l\PD{\bR{C}_l}{p_j}\right)\bigg|_{p_i=p_{i,\rm fid}}
	\,, \label{Eq:fisher}
\end{align}
where $l_\mathrm{max}$ is the maximum multipole included in the analysis, $\fsky$ is a sky fraction used for the analysis, $p_i$ are the parameters to be constrained, $p_{i,\rm fid}$ are the fiducial parameter values, and the covariance matrix of the observed $E$- and $B$ modes is given by
\begin{align}
    \bR{C}_l \equiv \Mat{ \hC_l^{EE} & \hC_l^{EB} \\ \hC_l^{EB} & \hC_l^{BB} }
	\,. 
\end{align}
The covariance matrix contains the total power spectra, $\hC^{EE}$, $\hC^{EB}$ and $\hC^{BB}$, including the lensed CMB, noise, and Galactic foregrounds after component separation. 
The $1\,\sigma$ constraint on $p_i$ is given by $\sigma(p_i)=(\{\bR{F}^{-1}\}_{ii})^{1/2}$. 

We consider two parameters, $p_1\equiv g\phi_{\rm in}/2$ and $p_2\equiv \alpha$, and set their fiducial values to be $p_{i,\rm fid}\equiv 0$. As the $E$- and $B$-mode spectra 
do not have a linear term of $p_i$ and their derivatives with respect to $p_i$ at $p_{i,\rm fid}= 0$ vanish, $\p\mathbf{C}_l/\p p_i|_{p_{i}= 0}$ contains only off-diagonal elements, $\p C^{EB}_l/\p p_i|_{p_{i}= 0}$. The Fisher matrix simplifies to \cite{Sherwin&Namikawa:2021}
\begin{align}
    \bR{F}_{ij} 
	&= \fsky\sum_{l=2}^{l_\mathrm{max}} \frac{2l+1}{\hC_l^{EE}\hC_l^{BB}}
	\PD{\hC^{EB}_l}{p_i}\PD{\hC^{EB}_l}{p_j}\bigg|_{p_i=0}
	\,. 
\end{align}

We assume a \textit{LiteBIRD}-like white noise ($2\mu$K-arcmin), angular resolution ($30$ arcmin), $\fsky=0.7$,
and the residual Galactic foregrounds obtained by Ref.~\cite{Errard:2016}. We set $l_\mathrm{max}=500$ because of the angular resolution. 

The impact of gravitational lensing on $C_l^{EB}$ would be negligible in the Fisher matrix. Lensing does not create $C_l^{EB}$ but only distorts small-scale polarization anisotropies. To see this, we first note that $C_l^{EB}$ is approximately given by \eq{Eq:EB:rei-rec}. As discussed in Ref.~\cite{Namikawa:2021:MC}, the two operators, lensing and birefringence, commute, and lensing replaces $C_l^{EE}$ in \eq{Eq:EB:rei-rec} with $C_l^{EE,\mathrm{lens}}-C_l^{BB,\mathrm{lens}}$. As $C_l^{BB,\mathrm{lens}}\ll C_l^{EE,\mathrm{lens}}$ at $l\ll 5000$ and $C_l^{EE}$ is modified by lensing only at high $l$, we ignore the gravitational lensing effect on $C_l^{EB}$.

\begin{figure}[t]
    \centering
    \includegraphics[width=\linewidth]{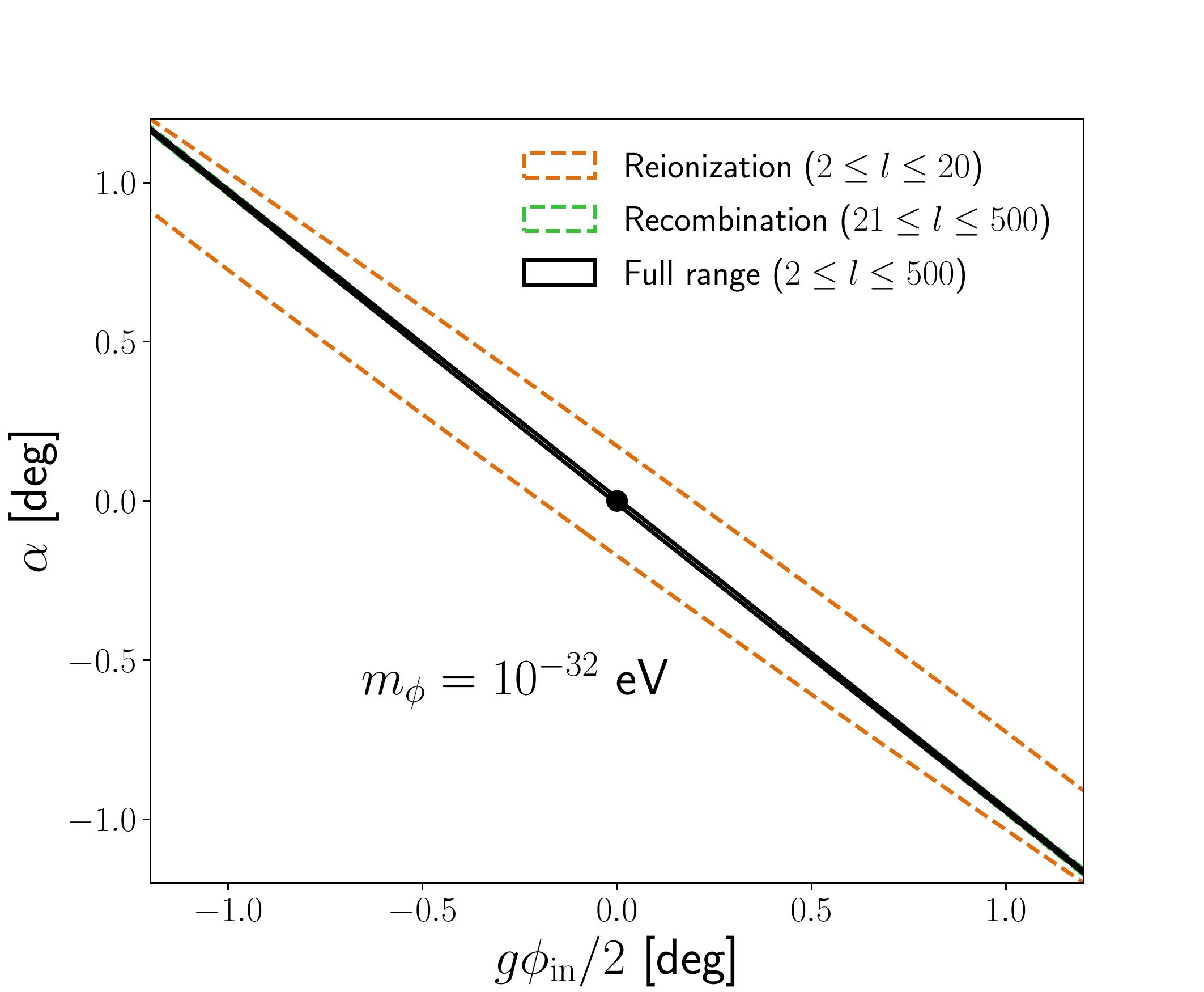}\\
    \includegraphics[width=\linewidth]{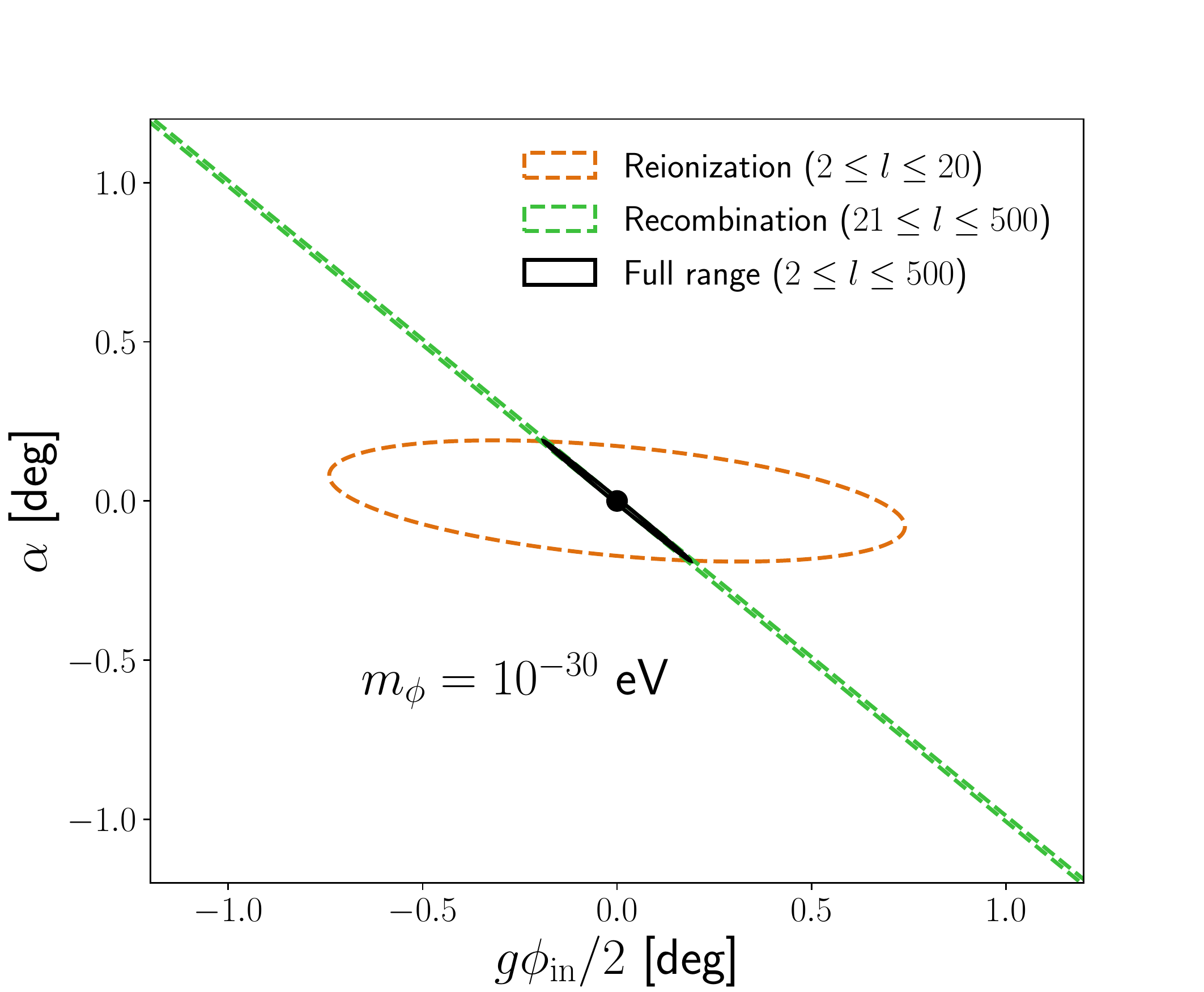}
    \caption{
    The expected $1\,\sigma$ error contours in the two-dimensional parameter space ($\alpha$ and $g\phi_\mathrm{in}/2$ in units of degrees). The axion masses are $m_\phi=10^{-32}$~eV (top) and $10^{-30}$~eV (bottom). We show the results from the full $l$ range (black), reionization bump ($2\le l\le 20$; orange dashed), and high multipoles ($21\le l\le 500$; green dot-dashed). We assume a \textit{LiteBIRD}-like experiment without delensing.
    }
    \label{fig_ellipse}
\end{figure}

Fig.~\ref{fig_ellipse} shows the expected error contours in the two-dimensional parameter space, $g\phi_{\rm in}/2$ and $\alpha$, for a given $m_\phi$. For $m_{\phi}=10^{-32}$\,eV the reionization bump in $C_l^{EB}$ is close to its maximum amplitude as explained in Sec.~\ref{sec:reionization}. In this case $C_l^{EB}$ becomes close to that of the miscalibration angle, and $g\phi_{\rm in}/2$ and $\alpha$ are strongly degenerate. For $m_{\phi}= 10^{-30}$~eV the reionization bump is suppressed and the degeneracy is reduced. We thus confirm and make more precise the result of Ref.~\cite{Sherwin&Namikawa:2021}.

\subsection{Constraining the axion mass}
\label{sec:CMB-S4}

Next, we consider joint constraints on $m_{\phi}$, $g\phi_{\rm in}/2$, and $\alpha$. Since $C_l^{EB}$ depends on $m_\phi$ nonlinearly, the Fisher matrix formalism, in which the errors are estimated from curvature of the posterior distribution around the fiducial value, does not provide accurate results. We thus use the likelihood analysis. We define $\chi^2$ as
\al{
    \chi^2(\mathbf{p}) = \fsky\sum_{l=l_{\rm min}}^{l_{\rm max}} (2l+1)\frac{[\hC^{EB}_l-C^{EB,\rm th}_l(\mathbf{p})]^2}{\hC^{EE}_l\hC^{BB}_l} 
    \,, 
}
where $C^{EB,\rm th}_l$ is a theoretical model for the $EB$ power spectrum and is given as the sum of the contributions from cosmic birefringence and $\alpha$.
For a given observed $\hC^{EB}_l$ we compute
$\chi^2$ for each parameter set, $\mathbf{p}=(m_\phi,g\phi_{\rm in}/2,\alpha)$, and obtain the posterior distribution, $P(\mathbf{p}|\hC^{EB})\propto\exp\left[-\chi^2(\mathbf{p})/2\right]$. 

We consider specifications similar to \textit{LiteBIRD}~\cite{LiteBIRD:2022}, Simons Observatory (SO; \cite{SimonsObservatory:2018koc}), and CMB-S4~\cite{CMB-S4:2016ple}. For SO we assume $l_{\rm min}=100$, $l_{\rm max}=2500$, $6\mu$K-arcmin white noise, $1$ arcmin Gaussian beam, $\fsky=0.4$, and no foregrounds. We also assume $30\%$ residual lensing-induced $B$ modes after delensing \cite{Namikawa:2021:SO}. For CMB-S4 we assume the same beam, $l_\mathrm{min}$, $l_\mathrm{max}$, and $\fsky$, but with $1\mu$K-arcmin noise and $10\%$ residual lensed $B$ modes~\cite{CMB-S4:2020:forecast}.

\begin{figure}[t]
\centering
\includegraphics[width=\linewidth]{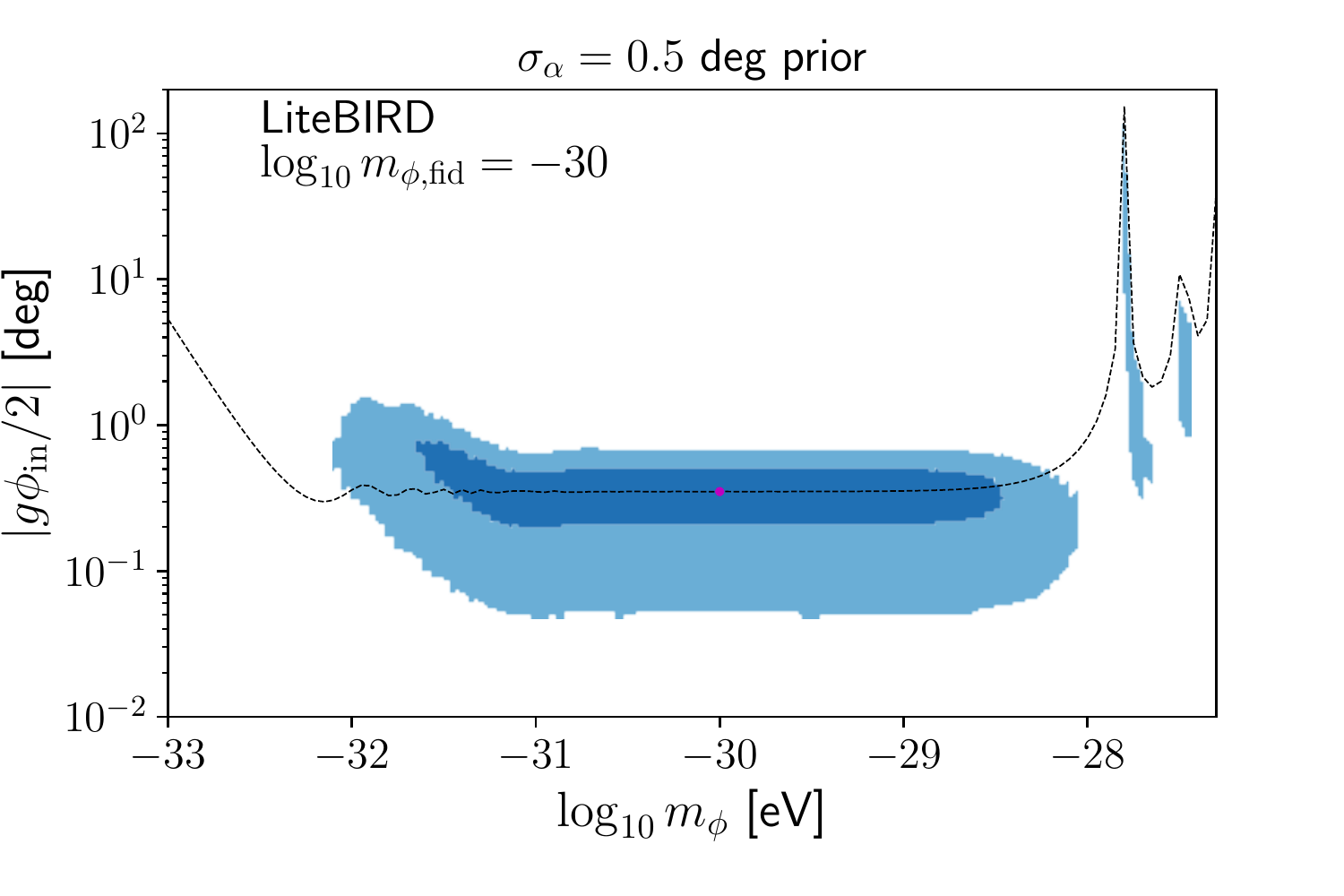}\\
\includegraphics[width=\linewidth]{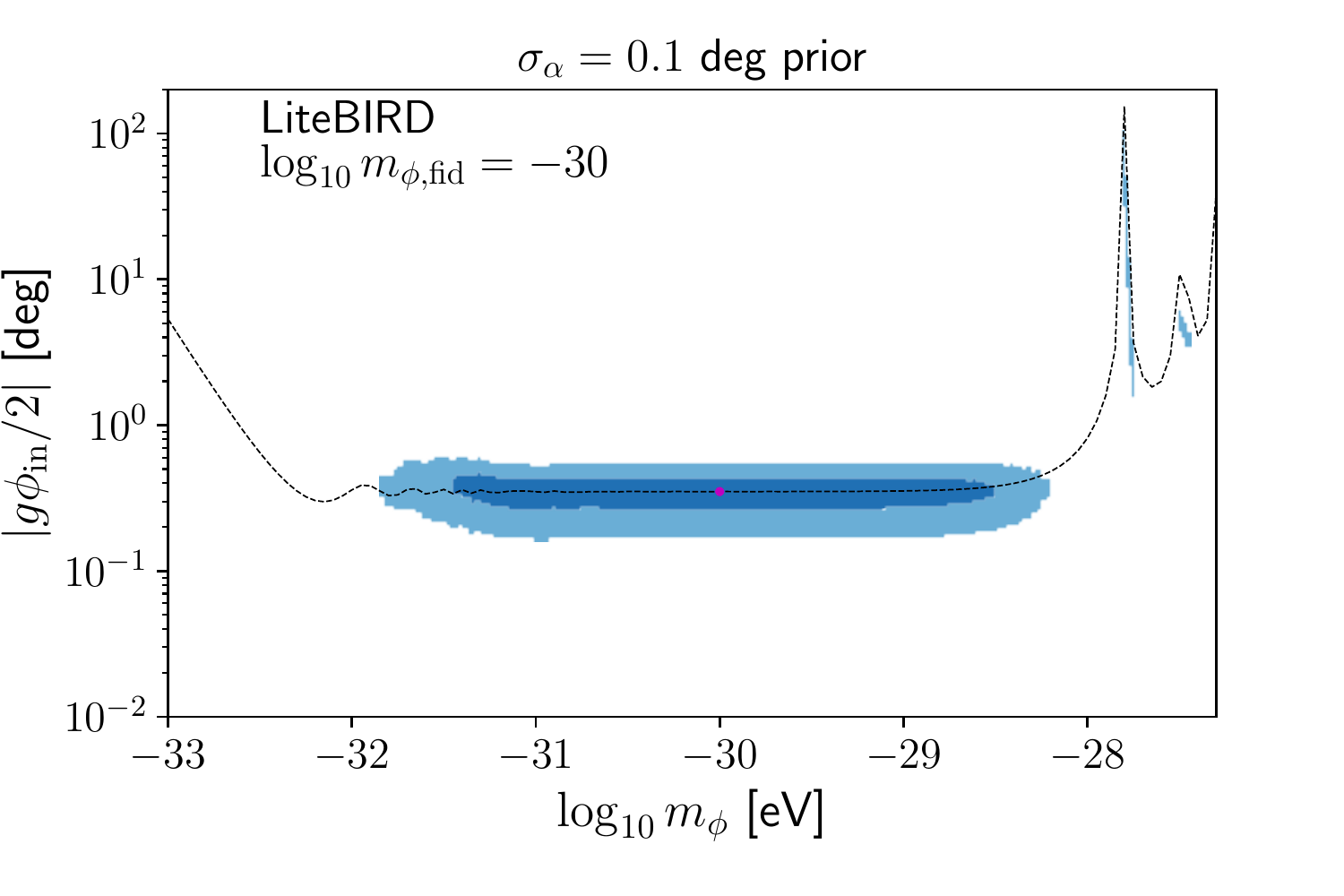}
\caption{
The expected 1 and 2\,$\sigma$ error contours on $\log_{10} m_{\phi}$ and $|g\phi_{\rm in}/2|$ (in units of degrees) for $m_{\phi,\mathrm{fid}}=10^{-30}$\,eV. 
The miscalibration angle is marginalized over using a Gaussian prior with $\sigma_\alpha=0.5$~deg (top) and 0.1~deg (bottom). The fiducial value of $g\phi_{\rm in}/2$ is chosen to give $\beta_{\rm rec}=0.35\,$deg (red dots). 
The black dashed lines show $|g\phi_{\rm in}/2|$ for $\beta_{\rm rec}=0.35\,$deg at each $m_\phi$. We assume a \textit{LiteBIRD}-like experiment.
}
\label{fig_forecast_3000_LB}
\end{figure}

Fig.~\ref{fig_forecast_3000_LB} shows the expected $1$ and $2$\,$\sigma$ contours on $\log_{10}m_{\phi}$ and $g\phi_{\rm in}/2$. We consider the \textit{LiteBIRD}-like experiment with specifications given in Sec.~\ref{sec:litebirdlike}. The fiducial axion mass is $m_{\phi,\mathrm{fid}}=10^{-30}\,$eV and $g\phi_{\rm in}/2$ is obtained from $\beta_{\rm rec}=0.35\,$deg~\cite{Minami:2020odp,Diego-Palazuelos:2022dsq,Eskilt:2022wav} via \eq{eq_beta_rec_approx}. The fiducial value of $\alpha$ is set to zero. We marginalize the posterior distribution over $\alpha$ 
using a prior distribution obtained from calibration of instruments. Specifically, we use a Gaussian prior,
$\exp[-\alpha^2/(2\sigma_\alpha^2)]$, with
$\sigma_\alpha=0.5$~deg (top) and 0.1~deg (bottom).
The former precision is achieved already for calibration of the current generation of CMB experiments~\cite{takahashi/etal:2010,PlanckIntXLIX,koopman:2018}, whereas the latter can be achieved by employing new calibration strategy~\cite{johnson/etal:2015,kaufman/keating/johnson:2016,nati/etal:2017,casas/etal:2021}.

The black dashed lines show the values of $|g\phi_{\rm in}/2|$ giving $\beta_{\rm rec}=0.35\,$deg for each $m_\phi$. The spikes in $m_\phi>10^{-28}$~eV occur when $C_l^{EB}$ at high $l$ becomes highly suppressed (see the blue line in Fig.~\ref{fig_ClEB_heavies_axion}). That is to say, we need a larger value of $|g\phi_\mathrm{in}/2|$ to compensate for suppression of $C_l^{EB}$ by a small value of $\braket{\phi}_\mathrm{LSS}$ [\eq{eq:phiLSS}].
 
We find that the prior on $\alpha$ tightens the constraint on the overall amplitude parameter ($g\phi_\mathrm{in}$) significantly, but $m_\phi$ can be constrained almost independently of the prior. We also find the same trend when removing the prior on $\alpha$ entirely. This is because the information on $m_\phi$ comes from the shape of $C_l^{EB}$. For $m_{\phi,\mathrm{fid}}=10^{-30}$\,eV the reionization bump is already at its minimum (see the blue line in Fig.~\ref{fig_betal_for_mass}). Therefore, the shape of $C_l^{EB}$ can tell us that $m_\phi$ is greater than $10^{-32}$~eV, but cannot tell how large it is until $m_\phi$ is so large that it affects the shape at high $l$. This explains an upper bound, $m_\phi\alt 10^{-28}$~eV. 

The ``islands'' of parameter space seen in $m_\phi> 10^{-28}$~eV are allowed because the peak locations of $C_l^{EB}$ for the respective $m_\phi$ in the islands happen to coincide with those for the fiducial mass of $10^{-30}$~eV, while the amplitudes are adjusted by varying $g\phi_\mathrm{in}$. The islands shrink when $g\phi_\mathrm{in}$ is constrained by a tighter prior on $\alpha$.

\begin{figure}[t]
\centering
\includegraphics[width=\linewidth]{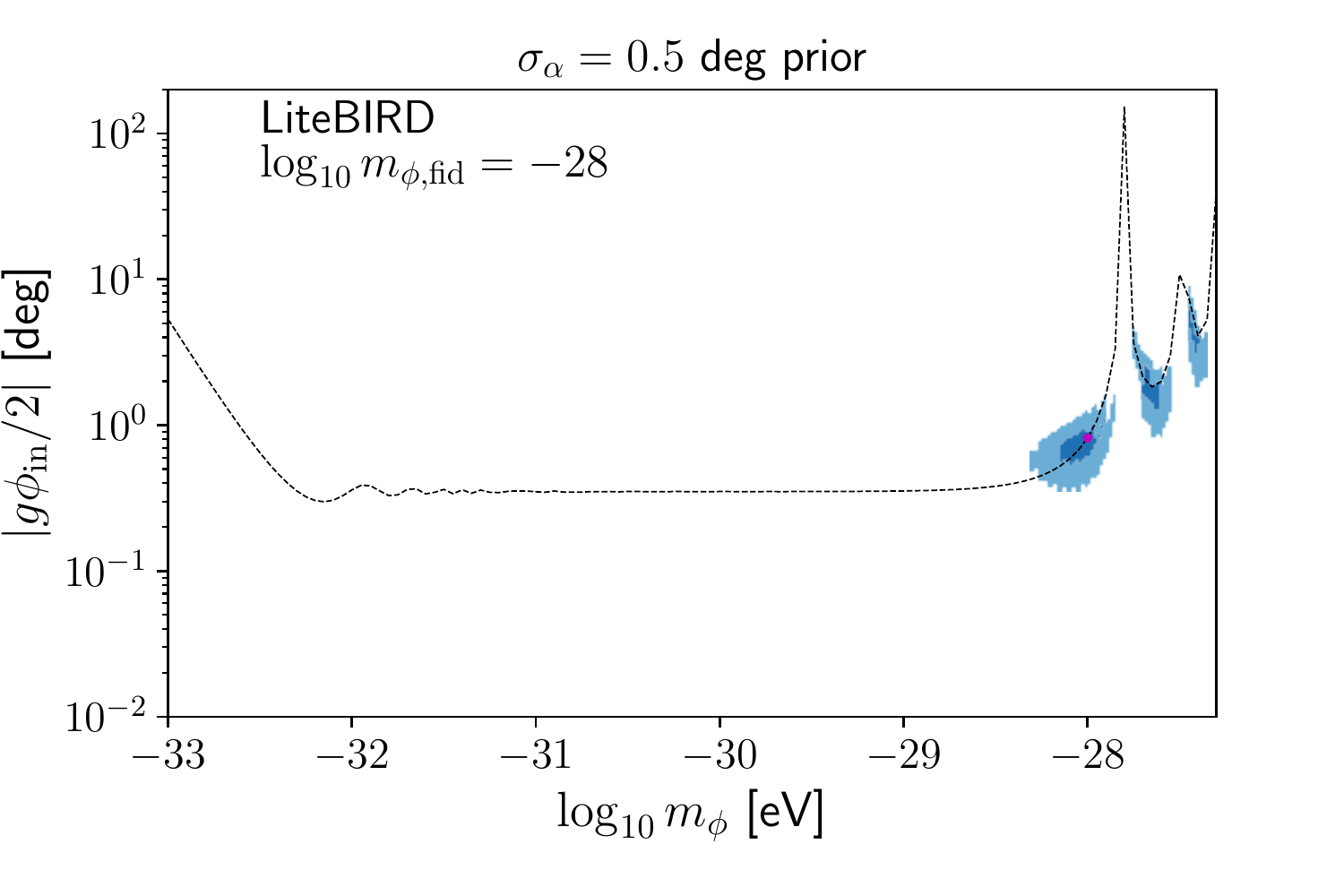}\\
\includegraphics[width=\linewidth]{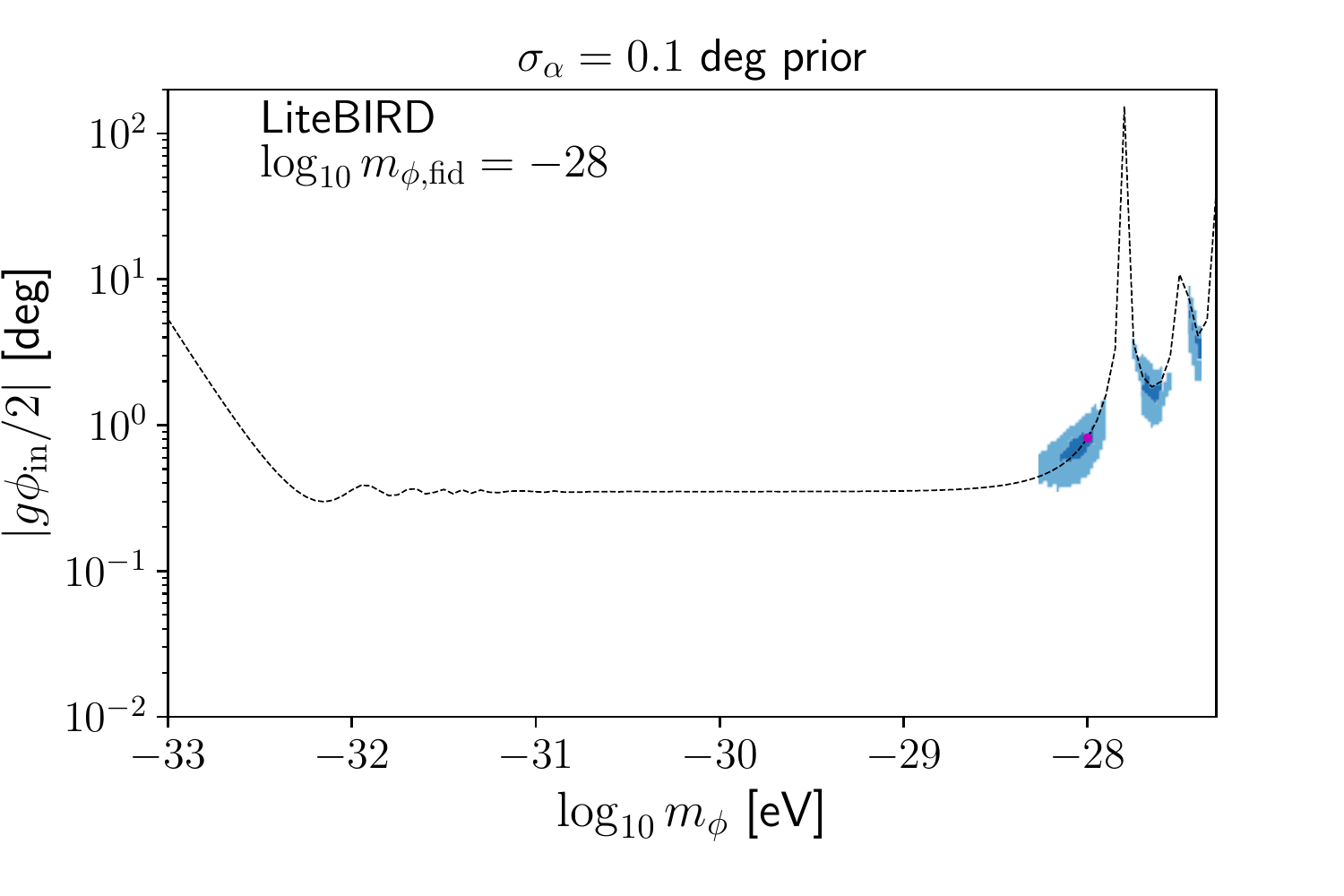}
\caption{
Same as Fig.~\ref{fig_forecast_3000_LB} but for $m_{\phi,\mathrm{fid}}=10^{-28}$\,eV. 
}
\label{fig_forecast_2800_LB}
\end{figure}

For $m_{\phi,\mathrm{fid}}=10^{-28}$\,eV the shape of $C_l^{EB}$ at high $l$ becomes quite different from that of $\alpha$, which enables us to determine $m_\phi$. 
However, the \textit{LiteBIRD}-like experiment cannot 
determine such a large $m_\phi$ accurately because of the limited angular resolution, giving only discrete islands of constraints (Fig.~\ref{fig_forecast_2800_LB}). Therefore, it gives effectively a lower bound for $m_\phi$ almost independently of $\sigma_\alpha$.

\begin{figure}[t]
\centering
\includegraphics[width=\linewidth]{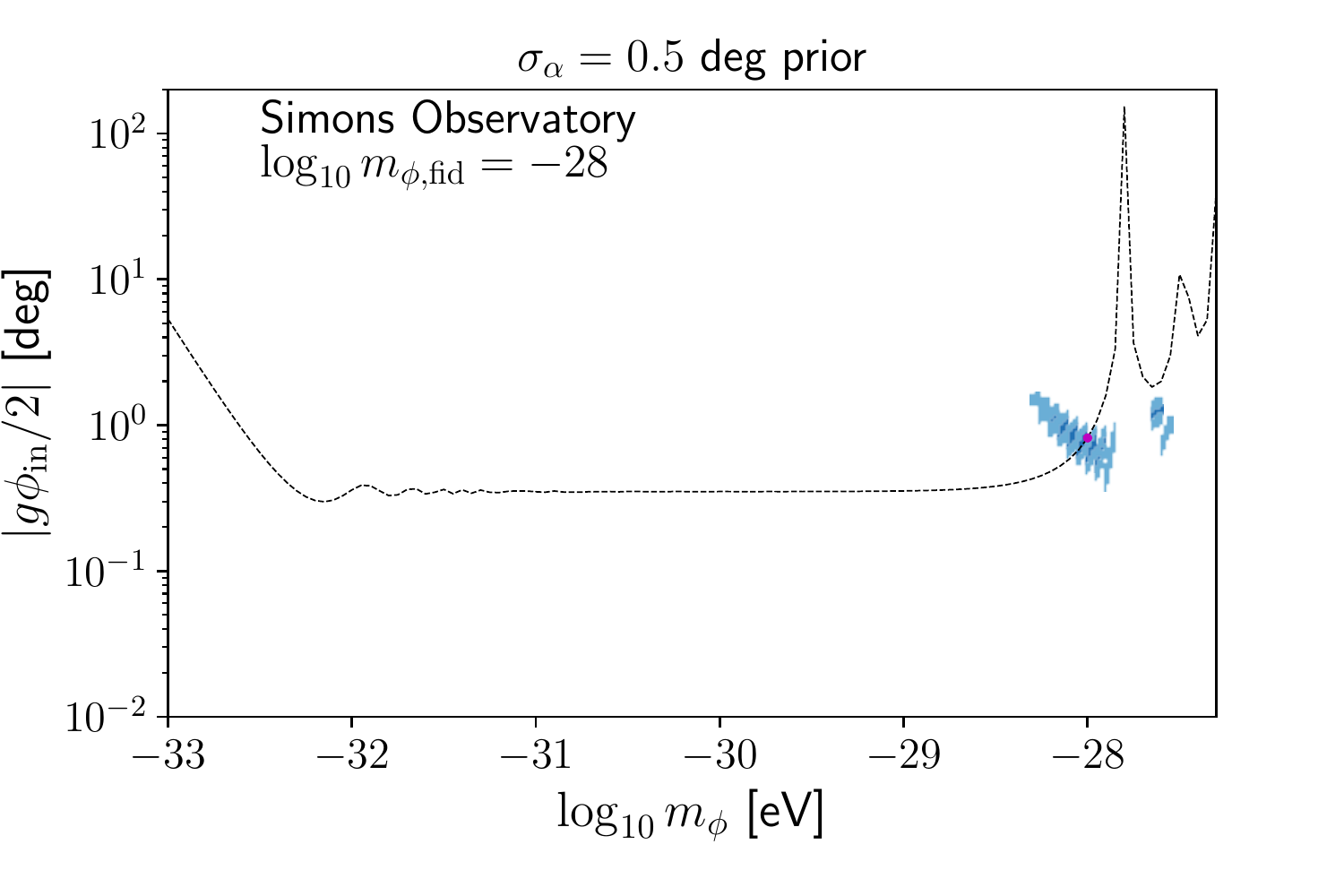}\\
\includegraphics[width=\linewidth]{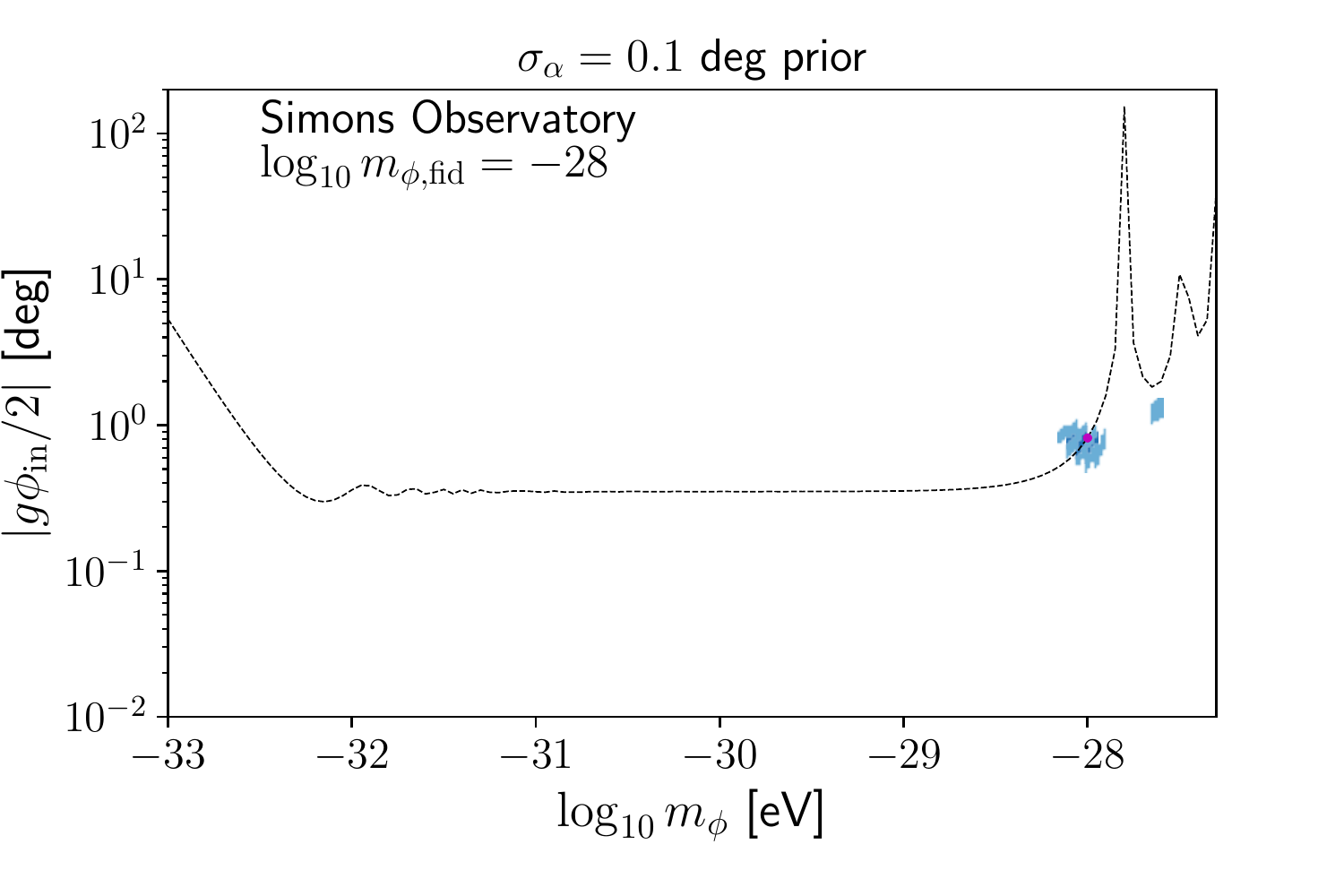}
\caption{
Same as Fig.~\ref{fig_forecast_2800_LB} but for an SO-like experiment.
}
\label{fig_forecast_2800_SO}
\end{figure}

\begin{figure}[t]
\centering
\includegraphics[width=\linewidth]{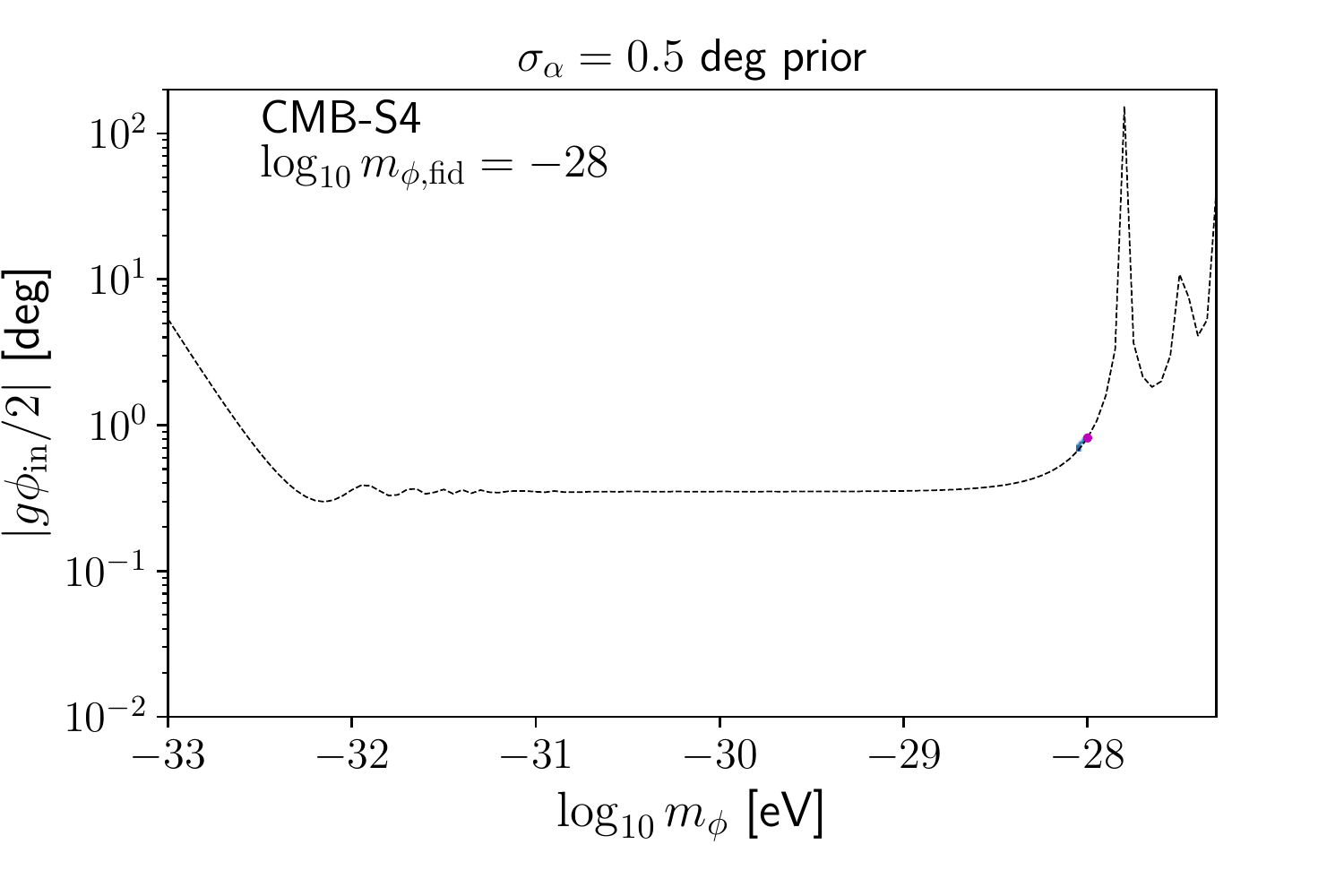}
\caption{
Same as the top panel of Fig.~\ref{fig_forecast_2800_SO} but for a CMB-S4-like experiment.
The result for $\sigma_\alpha=0.1$~deg is similar.
}
\label{fig_forecast_2800_S4}
\end{figure}

Ground-based experiments such as SO- and S4-like experiments have better sensitivity to large $m_\phi$. 
In Fig.~\ref{fig_forecast_2800_SO} we show the expected constraints for SO ($m_{\phi,\mathrm{fid}}=10^{-28}$\,eV).
The constraints tighten significantly compared to the \textit{LiteBIRD}-like case. Some degeneracy between $g\phi_{\rm in}$ and $m_{\phi}$ still exist for $\sigma_\alpha=0.5$~deg (top panel): the shift of peak locations can be absorbed partially by a combination of the rescaled $C_l^{EB}$ and $\alpha$.
The degeneracy is eliminated when a tighter prior is used (bottom panel). With CMB-S4 we can determine $m_\phi$ precisely, independent of $\alpha$ (see Fig.~\ref{fig_forecast_2800_S4}).

As SO and CMB-S4 cannot measure the reionization bump, they can only place an upper bound on $m_\phi$ if the fiducial mass is $10^{-30}\,$eV. Thus, \textit{LiteBIRD} and ground-based experiments are highly complementary. 

\section{Discussion}
\label{sec_discussion}

We have made some simplifying assumptions in our calculation of $C_l^{EB}$. 
First, we did not include the energy density of $\phi$ in the Friedmann equation explicitly.
This assumption can be justified to some extent. For $m_\phi\alt 10^{-33}$~eV, $\phi$ acts as dark energy and its energy density is included approximately as a cosmological constant in the Friedmann equation for $\Lambda$CDM cosmology.
For $10^{-32}~\mathrm{eV}\alt m_\phi\alt 10^{-25.5}~\mathrm{eV}$, $\phi$ acts as a small fraction of dark matter today with the density parameter $\Omega_\phi h^2 \alt 0.006$~\cite{Hlozek:2014lca}, which may be ignored for the current study. 
However, the change in shape of $C_l^{EB}$ at high $l$ is a subtle effect, which can be influenced quantitatively when $\Omega_\phi$ is included in the Friedmann equation.
We leave the full treatment for future work. 

Second, we did not vary cosmological parameters when calculating the expected future constraints on the axion parameters. This assumption can also be justified, as the effect of $\phi$ on $C_l^{EB}$ is distinct from the cosmological parameter dependence of the parity-even temperature and polarization power spectra. Nevertheless, there may still be some subtle correlation between the cosmological parameters and the axion parameters, which should be accounted for when the axion energy density is included in the Friedmann equation.

We now discuss possible future extensions of the calculation. We did not include inhomogeneity in $\phi$, which causes anisotropic polarization rotation~\cite{li/zhang:2008,pospelov/ritz/skordis:2009,kamionkowski:2009}. While there is no evidence for anisotropic birefringence~\cite{Contreras:2017sgi,Namikawa:2020ffr,SPT:2020cxx,Gruppuso:2020kfy}, it seems natural to expect discovery of anisotropy if the $3\sigma$ hint of isotropic birefringence~\cite{Minami:2020odp,Diego-Palazuelos:2022dsq,Eskilt:2022wav} is confirmed with higher statistical significance in future. Thus, incorporating anisotropic birefringence in the Boltzmann equation~\cite{Lee:2016jym,Capparelli:2019rtn} would be a natural next step.

We ignored the gravitational lensing effect on $C_l^{EB}$. The lensing would smear the acoustic peaks of $C_l^{EB}$ and enhance the small-scale power, but these effects would not be degenerate with cosmic birefringence. Nevertheless, for completeness, the impact of the lensing effect will be included in future work.

We have so far focused on cosmic birefringence by a single axion field, but it is entirely possible that cosmic birefringence is induced by multiple fields~\cite{Arvanitaki:2009fg,Obata:2021nql}. In this case, the time evolution of $\beta(\eta)$ becomes more interesting, which can be constrained by the tomographic approach. In this paper we considered two epochs, reionization and recombination, during which the polarization is efficiently generated. Other sources of polarization include the polarized Sunyaev-Zeldovich effect in clusters of galaxies, the so-called remote quadrupole \cite{Kamionkowski:1997:remote-quad,Portsmouth:2004:remote-quad,Deutsch:2017:remote-quad,Deutsch:2017:pSZ-tomography}. The polarization is generated after the epoch of reionization, and we can in principle use such large-scale polarization signals to probe $\beta(\eta)$ in a late-time Universe. 

\section{Conclusion}
\label{sec_conclusion}

In this paper, we solved the Boltzmann equation coupled with the EoM for an axionlike field $\phi$ to calculate the detailed shape of the $EB$ power spectrum of the CMB due to cosmic birefringence.  There are two critical axion masses: (1) $m_\phi\agt 10^{-32}$~eV, for which relative amplitudes of the reionization bump ($l\alt 10$) and the high-$l$ power spectrum are modified; and (2) $m_\phi\agt 10^{-28}$~eV, for which the evolution of $\phi$ during recombination yields complex features (such as a shift in the locations of acoustic peaks) at high $l$. Such a change in shape cannot be mimicked fully by the miscalibration angle $\alpha$, offering a powerful probe of $m_\phi$. In Ref.~\cite{Sherwin&Namikawa:2021}, this phenomenon was called ``cosmic birefringence tomography,'' as it allows us to measure the time evolution of $\phi$.

Probing the first critical mass requires a full-sky coverage by a satellite mission such as \textit{LiteBIRD}~\cite{LiteBIRD:2022}, whereas the second one can be probed by ground-based experiments~\cite{SimonsObservatory:2018koc,Moncelsi:2020ppj,CMB-S4:2016ple}. The important application of tomography is to distinguish whether $\phi$ is dark energy or (a fraction of) dark matter today. A convincing detection of the relative amplitude change of the low- and high-$l$ power spectrum by \textit{LiteBIRD} would rule out $\phi$ being dark energy. Ground-based experiments can constrain the value of $m_\phi$, especially at $\agt 10^{-28}$~eV, almost independently of $\alpha$. 
Together they can discover new physics and provide new scientific opportunities for CMB experiments~\cite{Komatsu:2022nvu}.

Finally, the $EB$ data at high $l$ from on-going ground-based CMB experiments such as Polarbear~\cite{Polarbear:2020lii}, Atacama Cosmology Telescope~\cite{ACT:2020frw}, and South Pole Telescope~\cite{SPT-3G:2021eoc} may already set interesting constraints on $m_\phi$.

\begin{acknowledgments}
We thank K. Murai, I. Obata, and M. Shiraishi for discussion and comments.
This work was supported in part by JSPS KAKENHI Grant No. JP19J21974 (H.N.), No. JP20H05850 (E.K.) and No. JP20H05859 (T.N. and E.K.), Advanced Leading Graduate Course for Photon Science (H.N.), the Deutsche Forschungsgemeinschaft (DFG, German Research Foundation) under Germany's Excellence Strategy - EXC-2094 - 390783311 (E.K.), and the European Union's Horizon 2020 research and innovation programme under the Marie Sk\l odowska-Curie grant agreement No.~101007633 (E.K.). The Kavli IPMU is supported by World Premier International Research Center Initiative (WPI), MEXT, Japan.
\end{acknowledgments}


\bibliographystyle{apsrev4-2}
\bibliography{Ref.bib}
\end{document}